\documentclass[aps,reprint,nofootinbib,onecolumn,superscriptaddress,amsmath,amssymb,byrevtex]{revtex4-2} %

\usepackage{graphicx}	
\usepackage{dcolumn}	
\usepackage{bm}		
\usepackage{multirow}  
\usepackage{makecell}
\usepackage{hyperref} \hypersetup{colorlinks=true, allcolors=blue}
\usepackage{amsmath}
\usepackage[dvipsnames]{xcolor}

\usepackage[commentmarkup=uwave,authormarkup=none, color=orange,defaultcolor=red]{changes} 

\newcommand\White{\color{White}}

\begin{document}


\title{Boundary-layer transition in the age of data: from a comprehensive dataset to fine-grained prediction}

\author{Wenhui Chang}
\altaffiliation{The two authors contributed equally.}
\affiliation{State Key Laboratory of Environment Characteristics and Effects for Near-space, Beijing Institute of Technology, Beijing 100081, China}

\author{Hongyuan Hu}
\altaffiliation{The two authors contributed equally.}
\affiliation{State Key Laboratory of Explosion Science and Safety Protection, Beijing Institute of Technology, Beijing 100081, China}

\author{Youcheng Xi}
\affiliation{School of Aerospace Engineering, Tsinghua University, Beijing 100084, China}

\author{Markus Kloker}
\affiliation{Institute of Aerodynamics and Gas Dynamics, University of Stuttgart, Pfaffenwaldring 21, D-70569 Stuttgart, Germany}

\author{Honghui Teng}
\email{hhteng@bit.edu.cn}
\affiliation{State Key Laboratory of Environment Characteristics and Effects for Near-space, Beijing Institute of Technology, Beijing 100081, China}

\author{Jie Ren}
\email{jie.ren@bit.edu.cn}
\affiliation{State Key Laboratory of Environment Characteristics and Effects for Near-space, Beijing Institute of Technology, Beijing 100081, China}

\date{\today}

\graphicspath{{Figures/}}


\begin{abstract}
The laminar-to-turbulent transition remains a fundamental and enduring challenge in fluid mechanics. Its complexity arises from the intrinsic nonlinearity and extreme sensitivity to external disturbances. This transition is critical in a wide range of applications, including aerospace, marine engineering, geophysical flows, and energy systems. While the governing physics can be well described by the Navier–Stokes equations, practical prediction efforts often fall short due to the lack of comprehensive models for perturbation initialization and turbulence generation in numerical simulations. To address the uncertainty introduced by unforeseeable environmental perturbations, we propose a fine-grained predictive framework that accurately predicts the transition location. The framework generates an extensive dataset using nonlinear parabolized stability equations (NPSE). NPSE simulations are performed over a wide range of randomly prescribed initial conditions for the generic zero-pressure-gradient flat-plate boundary-layer flow, resulting in a large dataset that captures the nonlinear evolution of instability waves across three canonical transition pathways (Type-K, -H, and -O). From a database of \(3000\) simulation cases, we extract diagnostic quantities (e.g., wall pressure signals and skin-friction coefficients) from each simulation to construct a feature set that links pre-transition flow characteristics to transition onset locations. Machine learning models are systematically evaluated, with ensemble methods—particularly XGBoost—demonstrating exceptional predictive accuracy (mean relative error of approximately 0.001). Compared to methods currently available (e.g., N-factor, transitional turbulence model), this approach accounts for the physical process and achieves transition prediction without relying on any empirical parameters. 
\end{abstract}


\maketitle

\section*{INTRODUCTION}\label{sec:S1}

Accurately predicting the laminar-to-turbulent transition is critical across a wide range of fluid flow applications, as turbulence can increase wall friction and heat flux by a factor of three to five compared to the laminar state \cite{fedorov2011transition,zhong2012direct,franko2013breakdown,chen2023constrained}, thereby directly affecting aerodynamic performance and thermal management. In high-speed aerodynamics, even minor inaccuracies in estimating the transition location for hypersonic vehicles can lead to catastrophic failures of thermal protection systems or substantial reductions in aerodynamic efficiency \cite{schneider2004hypersonic}. In blood flow, transition has been deemed a major contributing factor in the development of vascular diseases such as atherosclerosis, stenosis, and aneurysms, due to its strong association with the formation of atherosclerotic plaques \cite{saqr2020physiologic}. In turbomachinery operating under off-design conditions—such as dynamic stall or large-scale flow separation—laminar-to-turbulent transition significantly amplifies blade load fluctuations and increases fatigue risk \cite{nandi2017effects}. Despite its importance, the transition process remains challenging to observe and predict due to its extreme sensitivity to environmental conditions, geometric configurations, and underlying nonlinear dynamics \cite{annurev:/content/journals/10.1146/annurev.fluid.34.082701.161921,bertin2006critical,avila2011onset,caulfield2021layering,wu2023new}.

In pursuit of a practical solution for transition prediction, classical approaches based on linear stability theory (LST) have been widely used to analyze the growth rates of flow perturbations \cite{gaster1975theoretical, mack1984boundary, schmidstability}. The N-factor, obtained by integrating the disturbance growth rate along the streamwise coordinate, offers a crude overall characterization of disturbance amplification. This method, often combined with empirical calibration, remains in use today. However, a critical limitation of the N-factor method is that it neglects the real receptivity stage—the process by which internal or external perturbations interact with the boundary-layer and establish the initial conditions for instability growth. Recent efforts, such as the integration of bi-orthogonal decomposition with LST, aim to enhance receptivity predictions by enabling the extraction of mode amplitudes in complex, multimodal environments \cite{zou2024new}. In parallel, the development of parabolized stability equations (PSE) in the early 1990s marked a major breakthrough in addressing non-parallel and weakly nonlinear effects in boundary-layer flows \cite{bertolotti1992linear, chang1991compressible, herbert1997parabolized}. The PSE framework involves decomposing disturbances into rapidly oscillating wave functions and slowly varying envelope (shape) functions, thereby transforming the equations into a parabolic form. This formulation allows computationally efficient downstream-marching simulations of disturbance evolution up to breakdown onset, typically marked by the rapid rise of the skin-friction coefficient ($C_f$) curve \cite{joslin1993spatial}.

Another approach is direct numerical simulation (DNS) which solves the full Navier-Stokes equations without artificial modeling, enabling a comprehensive analysis of fluid flow while inherently capturing the receptivity stage. This capability is particularly critical in scenarios where experimental diagnostics are constrained \cite{hogberg2003linear,moin1998direct}. The seminal work by \citet{fasel1990numerical} demonstrated DNS's ability to simulate the nonlinear evolution of Tollmien-Schlichting (T-S) waves, reproducing the formation of $\Lambda$-vortices and quantifying the critical amplitude for secondary instabilities \cite{kloker1990numerical,rist1995direct}. Subsequent DNS investigations have further clarified the dynamics of boundary-layer transition, with \citet{andersson2001breakdown} identifying transient growth thresholds that drive streak formation under free-stream turbulence. Moreover, \citet{wu2009direct} provided a detailed numerical reconstruction of bypass transition, emphasizing spatiotemporal interactions among instability modes. Integrated experimental-DNS methodologies, such as those developed by \citet{saric2019experiments}, have elucidated how surface roughness modulates both receptivity and instability development, thereby offering deeper insights into the transition process. However, the high computational cost of DNS makes it impractical for many engineering applications.

From the perspective of the current era of big data, although machine learning has emerged as a transformative paradigm across numerous scientific and engineering fields, its application to transition prediction remains preliminary and largely exploratory. Several notable examples have demonstrated early successes. Hybrid convolutional neural networks (CNNs) have been used to encode boundary-layer velocity profiles into latent features for predicting local instability amplification, eliminating the need for eigenvalue solvers \cite{zafar2020convolutional}. Recurrent neural networks (RNNs) extend this by learning N-factor envelopes directly from mean flows \cite{zafar2021recurrent}. To address the challenges of turbulence modeling at high Reynolds numbers, physics-informed neural networks (PINNs) integrate Navier--Stokes residuals into the training objective, combining sparse data with physical constraints \cite{ghosh2023rans}. When coupled with Reynolds-Averaged Navier--Stokes (RANS) models, PINNs offer fast, differentiable surrogates across diverse geometries and can reconstruct near-wall velocity fields using limited off-wall measurements \cite{sekar2021accurate}. For unstructured domains, graph neural networks (GNNs) support flow reconstruction where grid-based methods fail, thus broadening machine learning (ML) applicability to complex geometries \cite{kashefi2021point}. Operator-learning methods such as Fourier Neural Operators (FNOs) further enhance generalization across parameterized systems by learning mesh-independent mappings \cite{li2021fourier}. In extreme regimes like hypersonic flows with thermochemical nonequilibrium, deep operator networks (DeepONets) and their variants (e.g., DeepM\&Mnet) achieve rapid, accurate inference of coupled fields—velocity, temperature, and species—outperforming traditional solvers by several orders of magnitude in speed \cite{mao2020deepm}.

In summary, the current paradigm for transition prediction faces several critical challenges. 
First, classical stability theory lacks a comprehensive framework for modeling the receptivity stage, which can manifest in diverse forms depending on the nature and source of external perturbations. These perturbations are inherently varied and difficult to parameterize.
Second, nonlinear mechanisms remain inadequately modeled and quantified due to the complex interactions among disturbances and the multiple pathways through which transition may occur \cite{franko2013breakdown, wagnild2012vibrational, lozano2018modeling}.
Third, although DNS offers high-fidelity insights, their application to realistic configurations—especially those involving a range of possible external perturbations—is severely constrained by the prohibitive computational cost \cite{spalart2000strategies}. Additionally, experimental approaches often suffer from limited spatial resolution and inconsistencies between wind tunnel conditions and actual flight environments \cite{nakagawa2023dns,duan2020simulation}.
To overcome these limitations, this work proposes a synergistic data-driven machine learning framework that leverages a large dataset generated using Nonlinear Parabolized Stability Equations(NPSE). The dataset incorporates various initial conditions to capture a broad spectrum of receptivity scenarios. Flow features—designed to emulate wall pressure measurements—are extracted and labeled with their corresponding transition onset locations. The constructed framework is incrementally extensible, with the ultimate goal of improving transition prediction accuracy for complex and realistic flow configurations. \cite{vlachas2022multiscale}

\section*{RESULTS}\label{S2}
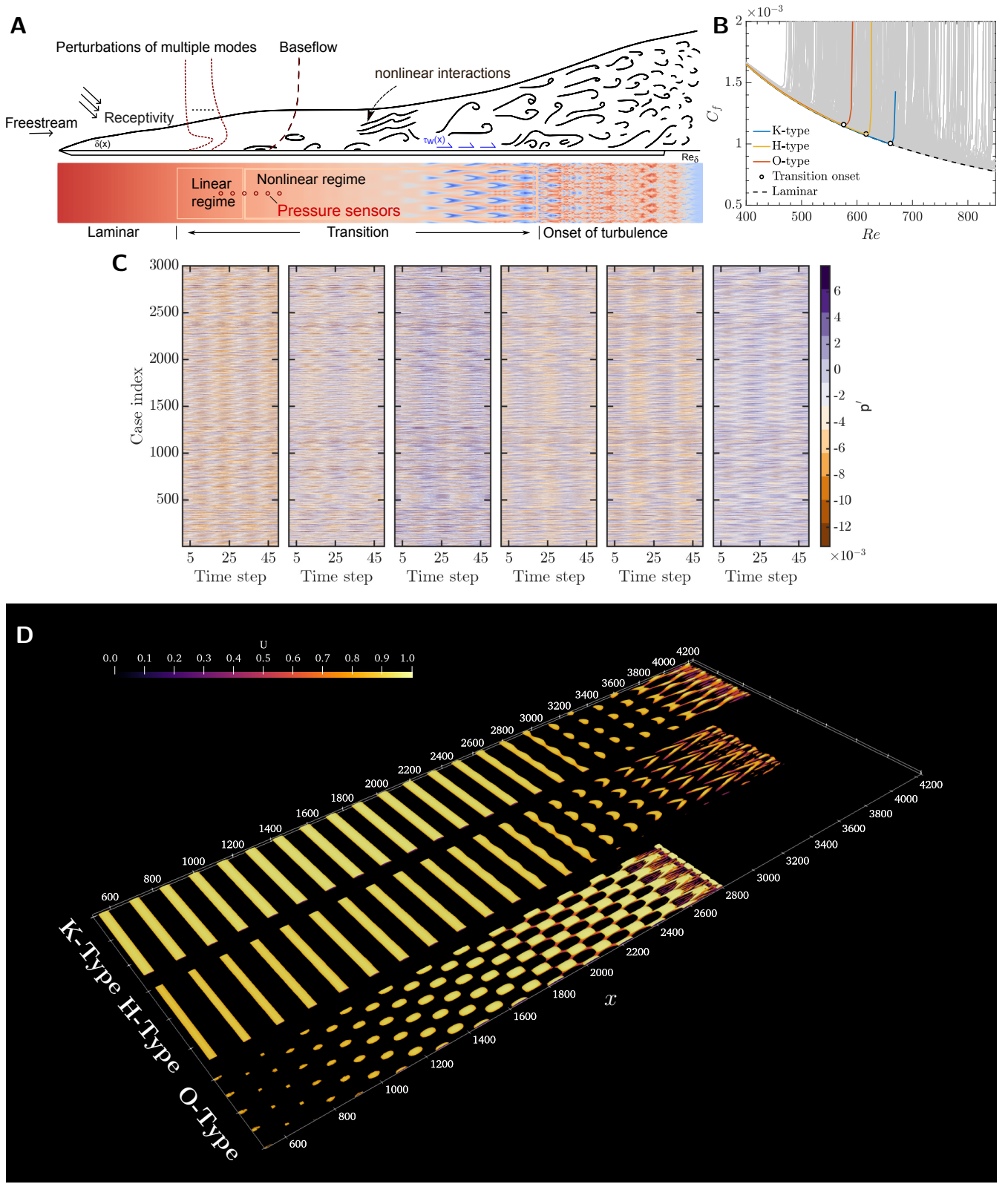
\begin{figure}
\centering
\includegraphics[width=0.7\textwidth]{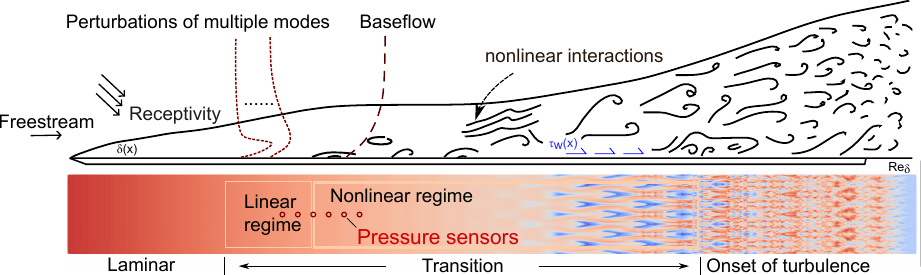} \put(-355, 105){\large \bfseries\sffamily A}
\includegraphics[width=0.29\textwidth]{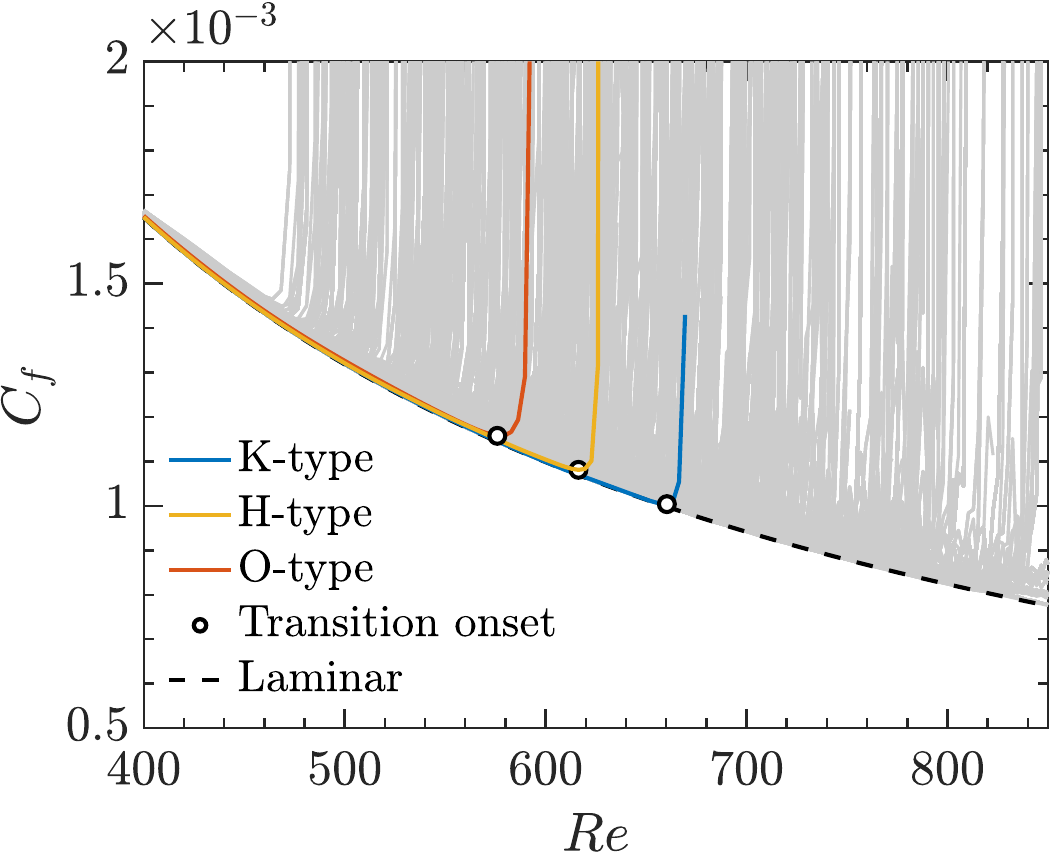} \put(-145, 105){\large \bfseries\sffamily B} \\[2mm]
\includegraphics[width=0.75\textwidth]{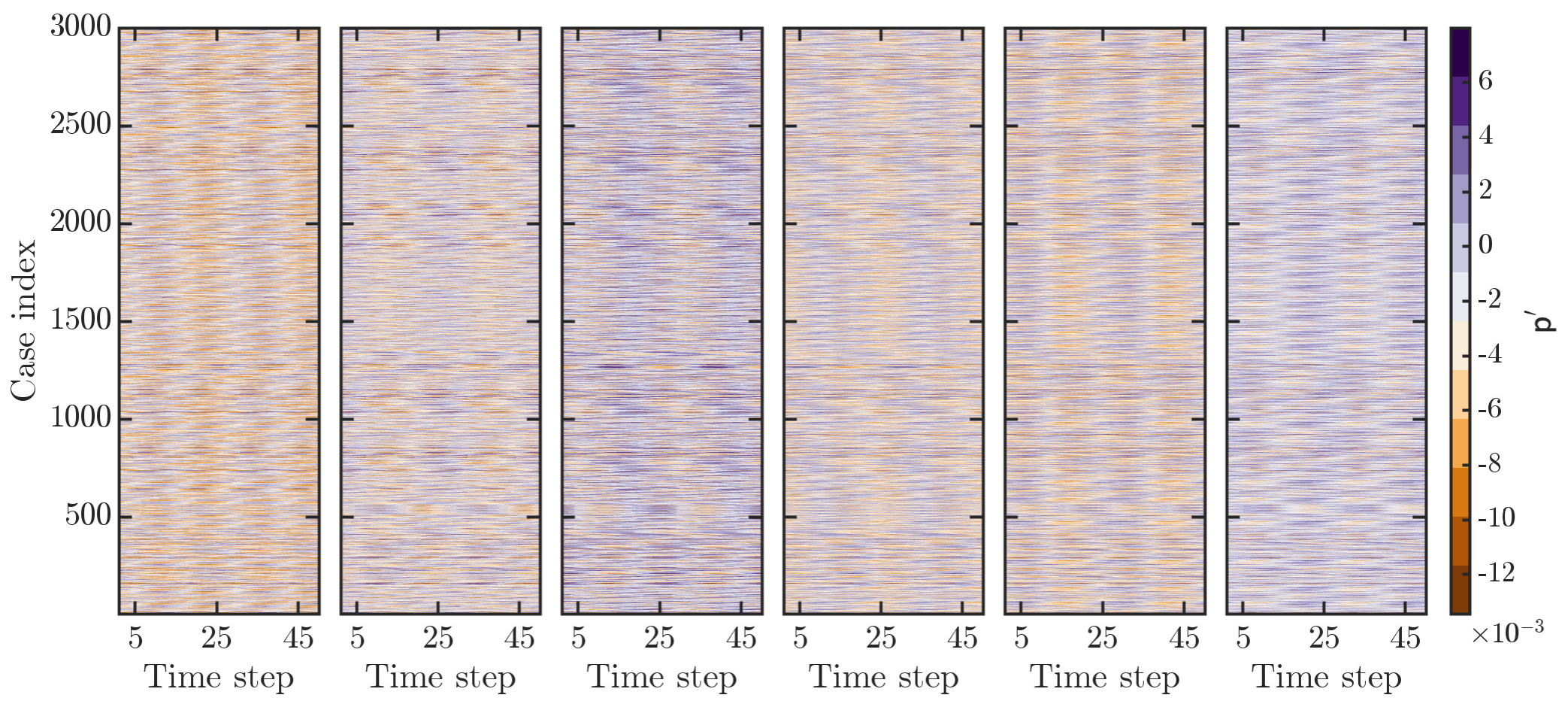}
\put(-390, 163){\large \bfseries\sffamily C}
\\[2mm]
\includegraphics[width=0.99\textwidth]{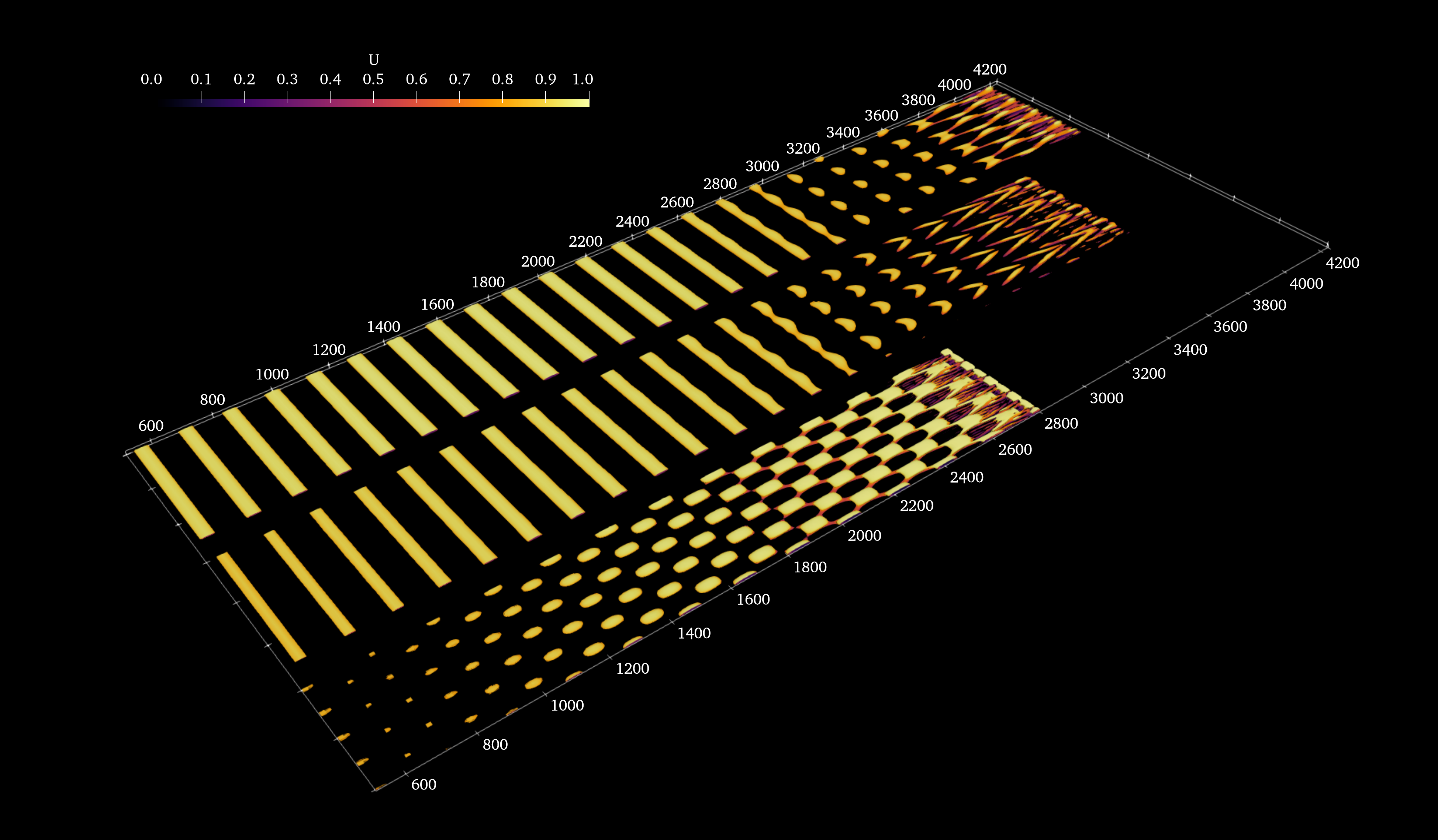} 
\put(-500, 275) {\White \large \bfseries\sffamily D} 
\put(-480, 130) {\rotatebox{-51}{\White \large \bfseries K-Type}}
\put(-450, 90) {\rotatebox{-51}{\White \large \bfseries H-Type}}
\put(-420, 45) {\rotatebox{-51}{\White \large \bfseries O-Type}}
\put(-200, 90) {\rotatebox{0}{\White \large $x$}}
\caption{
\textbf{\sffamily Problem Definition, methods and data.} 
\textbf{\sffamily A} Boundary-layer transition process Schematic,Freestream excites perturbations via receptivity, triggering nonlinear interactions between baseflow and perturbations that induce transition.
\textbf{\sffamily B} Wall friction coefficient \(C_f\) versus \(Re\), including laminar Blasius solution (hollow circle) and three dominant pathways (K-type, O-type, H-type). Red circles denote onset locations. $Re=\rho_\infty U_\infty\delta/\mu_\infty$, $\delta=\sqrt{\mu_\infty x/(\rho_\infty U_\infty)}$.
\textbf{\sffamily C} Perturbation pressure  at $x = 420,\ 473,\ 526,\ 579,\ 632,\ \text{and}\ 685$. Vertical axis shows all simulation samples.
\textbf{\sffamily D} Isosurfaces of the Q-criterion ($Q = 0.00001$) illustrating three canonical transition scenarios: K-type, H-type, and O-type. Each visualizes the coherent vortical structures characteristic of the respective transition mechanism.
}
\label{fig1}
\end{figure}

\subsubsection*{Flow features and dataset}
Based on 3000 high-fidelity simulations generated with the NPSE framework, a comprehensive dataset has been established that incorporates various receptivity scenarios. The simulations provide quick and reliable data up to the rise of the \(C_f\) curve, which defines the onset of flow transition. The fidelity of the NPSE in resolving transitional flow has been discussed in previous studies \cite{joslin1993spatial,lozano2018modeling}. A validation of the NPSE against DNS is provided in Appendix \ref{secA1}. Double-spectral notation $(m, n)$ has been used to denote a mode with frequency $m\omega_0$ and spanwise wavenumber $n\beta_0$ where $\omega_0$ and $\beta_0$ are the fundamental frequency and wavenumber. The dataset includes a range of initial conditions that cover three major natural transition types:

\begin{itemize}
\item {Klebanoff-type \cite{klebanoff1959evolution} transition (K-type):} 
Transition is initiated by the linear amplification of two-dimensional Tollmien–Schlichting (T-S) waves (mode $(2,0)$). These waves grow exponentially downstream, and once they reach a sufficient amplitude, they interact with three-dimensional secondary instabilities of the same frequency (mode $(2,\pm 1)$), known as the fundamental resonance. These disturbances further develop, contributing to the formation of aligned periodic \(\Lambda\)-vortices, which eventually break down and coalesce into localized turbulent spots.

\item {Herbert-type \cite{herbert1988secondary}  transition (H-type):}  
The dominant three-dimensional disturbances occur at half the frequency of the fundamental T-S wave (modes $(2,0)$ (primary) and $(1,\pm 1)$ (secondary)). This subharmonic resonance leads to staggered \(\Lambda\)-vortices, which are the hallmark of H-type transition. Theoretically, the secondary stage starts at somewhat lower T-S-wave amplitude level than in K-type transition, but since no significant steady mode (0,1) is involved in H-type as with K-type, the latter often prevails due to higher effective receptivity.

\item {Oblique-wave transition (O-type):}  
It arises from the interaction between a pair of disturbances with the same frequency but opposite wave angles (mode $(1,1)$ and $(1,-1)$). These interactions generate alternating high- and low-speed streaks in the boundary-layer \cite{schlichting2000fundamentals} prior to breakdown. O-type breakdown is especially important with point-like excitation sources or at higher subsonic or supersonic flow speeds.
\end{itemize}

The dataset includes three canonical transition scenarios: K-type, H-type, and O-type. The present study focuses on canonical two-dimensional boundary-layer flow under zero pressure gradient with a low turbulence level, where there is no breakdown of low-frequency streaks. As shown in Figure~\ref{fig1}, $C_f$ profiles are used to identify the onset of transition. The results indicate that the transition occurs within the range $Re = 480$ to $Re = 850$, covering approximately 82\% of the computational domain ($Re = 400-850$); the $Re$ number uses the local Blasius length scale which is 59\% of the local boundary-layer displacement thickness. To emulate wall-pressure measurements, time-resolved pressure signals were extracted at six streamwise locations ($x = 420,\ 473,\ 526,\ 579,\ 632,\ \text{and}\ 685$). These correspond to Reynolds numbers of $Re = 400,\ 417,\ 439,\ 461,\ 482,\ \text{and}\ 502$, respectively. These signals exhibit strong periodicity, suggesting the presence of a dominant mode in the boundary-layer. At each position, 50 consecutive time steps were recorded, forming the basic input for subsequent machine learning analysis. The sampling window spans 2--3 full cycles of the primary disturbance, with the fundamental period given by
\begin{equation}
    T_0 = \frac{2\pi}{\omega_0},
\end{equation}
ensuring that the essential nonlinear features of the flow are fully captured.

\subsubsection*{Transition Prediction}\label{S2c}
\begin{figure}
\centering
\includegraphics[scale=0.6]{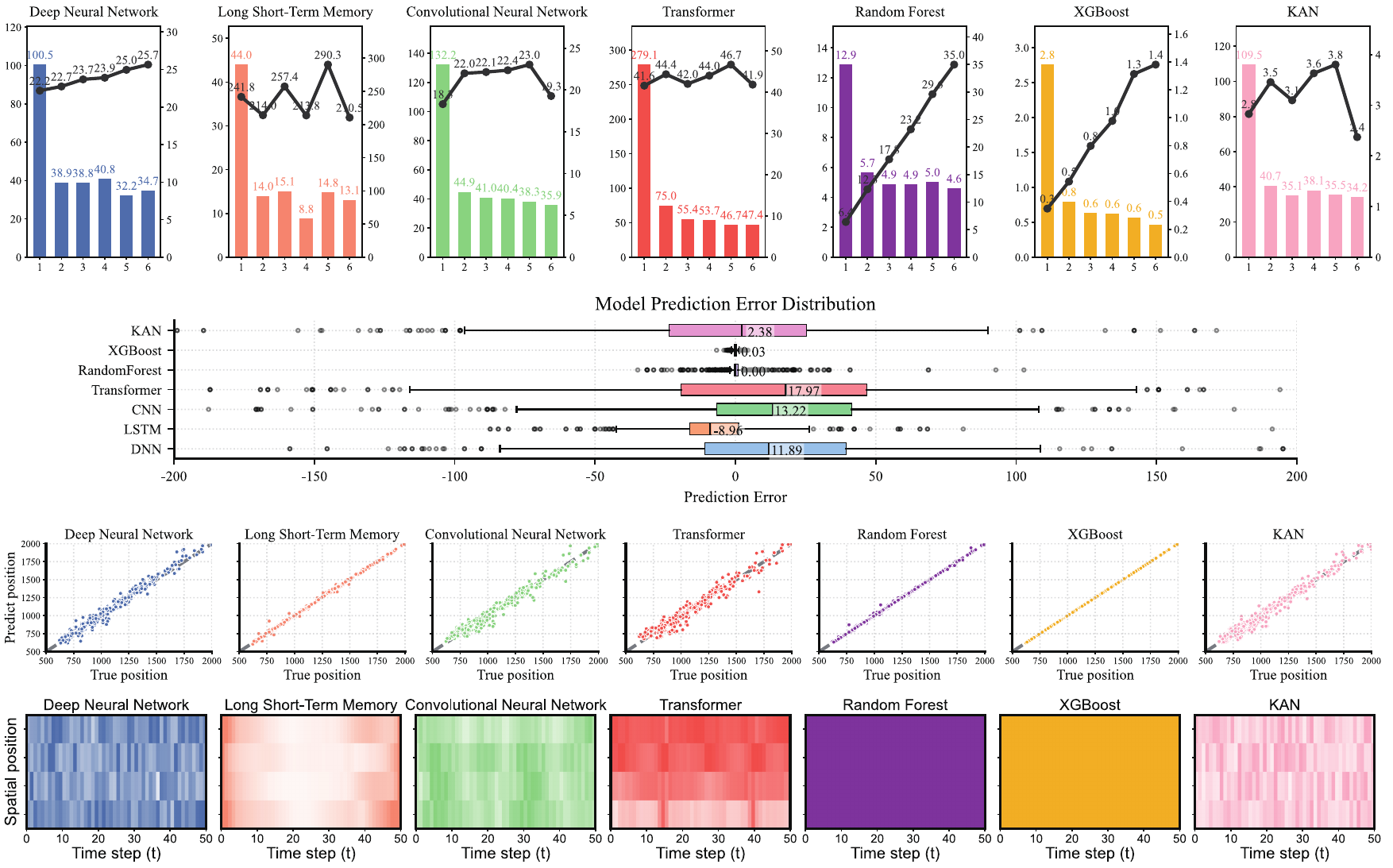}\put(-510,308){\large \bfseries\sffamily A}\put(-510, 190){\large \bfseries\sffamily B}\put(-510, 133){\large \bfseries\sffamily C}\put(-510, 67){\large \bfseries\sffamily D}
\caption{
\textbf{\sffamily Comparative assessment of predictive models for transition location estimation.}
\textbf{\sffamily A} Influence of the number of wall pressure sampling points on the predictive performance and computational cost of seven models. The horizontal axis represents the number of sampling points used. Bars show  MAE, while the overlaid line indicates training time.
\textbf{\sffamily B} Box plots depicting the distribution of prediction errors across all evaluated models. The horizontal axis represents the prediction error, and the vertical axis lists the model types. Numerical annotations indicate the median error for each model.
\textbf{\sffamily C} Scatter plots comparing predicted versus true transition locations on the validation dataset for the seven models. The horizontal axis represents the ground truth transition location, while the vertical axis shows the corresponding model predictions.
\textbf{\sffamily D} Feature importance analysis for the seven models. The x-axis represents the 50 time steps, and the y-axis represents the four spatial positions. Color intensity indicates the magnitude of feature importance, with darker shades corresponding to higher importance values.
}
\label{fig:performance}
\end{figure}
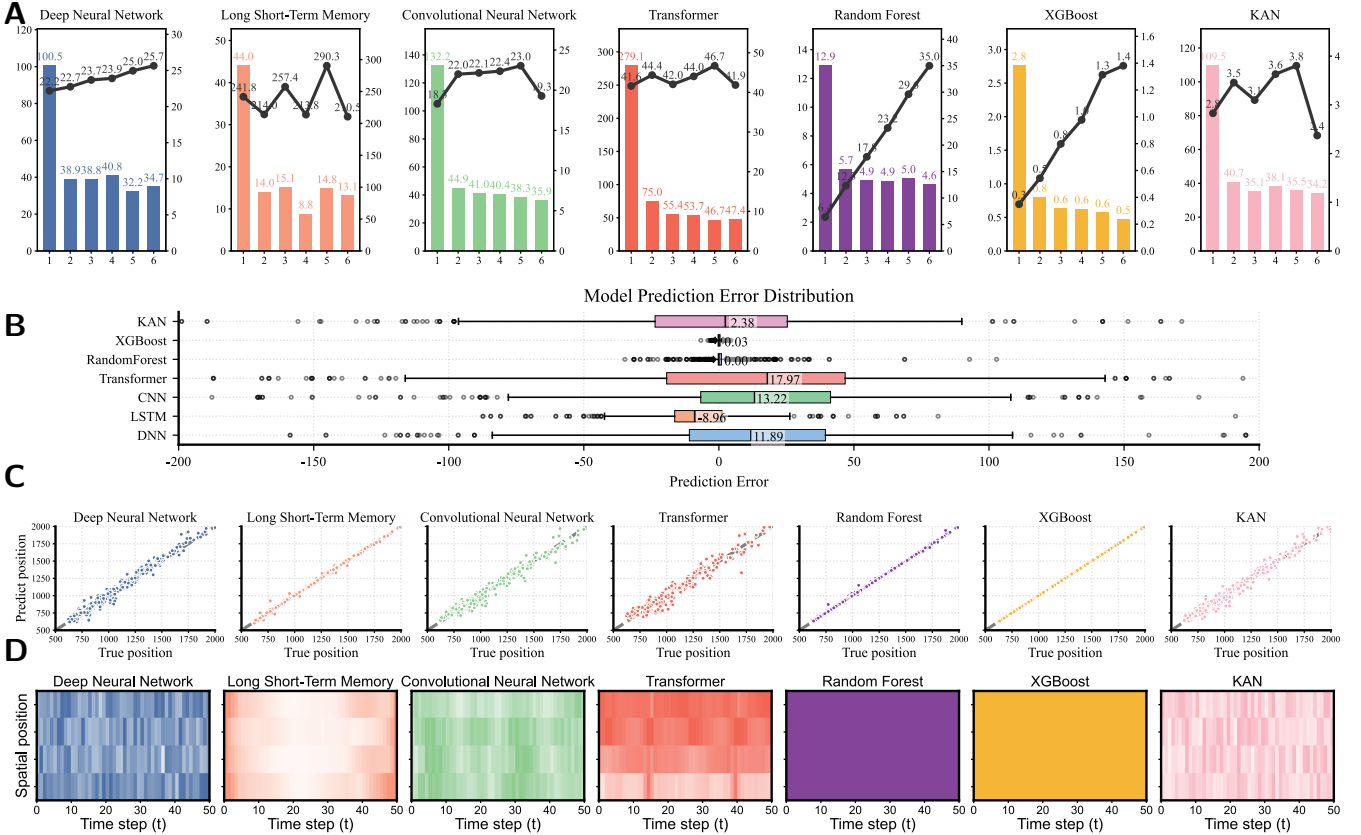

To determine the optimal number of wall pressure detection points, a series of numerical experiments were conducted, as shown in Figure \ref{fig:performance}a. Considering both the predictive performance and computational cost of the models, and in view of engineering practicalities, four streamwise wall-pressure sampling points were selected for subsequent transition prediction analysis.

The predictive performance of the investigated models was evaluated using standard regression metrics, including mean squared error (MSE), mean absolute error (MAE), mean relative error (MRE) and the \(R^2\) score
\begin{equation}
R^2 = 1 - \dfrac{\sum_{i=1}^{n} (x_i - \hat{x}_i)^2}{\sum_{i=1}^{n} (x_i - \bar{x})^2},
\end{equation}
where \(x_i\) is the actual value, \(\hat{x}_i\) is the predicted value, and \(\bar{x}\) is the mean of the actual values). As summarized in Table~\ref{tab:model_performance}, the seven models exhibit notable differences in their ability to predict transition onset locations. Ensemble methods, especially XGBoost and Random Forest, show superior accuracy and computational efficiency. In contrast, some deep learning architectures, particularly the transformer model, demonstrated performance anomalies that deviated from theoretical expectations.

\begin{table}
\centering
\caption{Comparison of evaluation metrics and computational resource consumption for seven machine learning algorithms.}
\label{tab:model_performance}
\begin{tabular}{lrrrrlc}  
\hline
\textbf{Model} & \textbf{Training Time (s)} & \textbf{MSE} & \textbf{MAE} & \textbf{MRE} & \textbf{R\textsuperscript{2}} & \textbf{Size (MB)} \\
\hline
DNN         		& 23.9  & 2837.94   & 40.79   & 0.037 & 0.9723 & 0.7   \\
LSTM        	& 213.8  & 1035.13   & 8.78  & 0.011  & 0.9980 & 1.4  \\
CNN         		& 22.3  & 2853.11   & 40.43  & 0.036  & 0.9708 & 0.5  \\
Transformer 	& 44.0  & 10346.86  & 53.69  & 0.047  & 0.9533 & 0.3  \\
Random Forest	& 23.2  & 428.92    & 4.86  & 0.004  & 0.9989 & 25   \\
XGBoost     	& 0.9   & 27.67     & 0.62  & 0.001  & 0.9999 & 25   \\
KAN         		& 3.6   & 7312.84   & 38.06  & 0.037  & 0.9741 & 0.4 \\
\hline
\end{tabular}
\end{table}

Figure~\ref{fig:performance}(b,c) presents a detailed comparison of predictive accuracy and error distributions across the models. Extreme gradient boosting (XGBoost) achieved the best overall performance, with an MRE of 0.001 and an MSE of 27.67, corresponding to an \(R^2\) value of 0.9999—close to the theoretical optimum. These results demonstrate XGBoost's exceptional ability to capture the nonlinear relationships between wall pressure time series features and transition onset locations. Random Forest ranked second, with an MRE of 0.004, approximately 4 times higher than XGBoost, indicating relatively lower predictive accuracy. However, the error distribution analysis revealed that Random Forest exhibited greater stability, with a lower standard deviation and a narrower interquartile range (IQR), suggesting it is more robust compared to other models. Among the deep learning models, Long Short Term Memory Network(LSTM) (MRE = 0.011) and CNN (MRE = 0.036) outperformed the deep neural network (DNN) and Kolmogorov-Arnold network(KAN) models, though they still lagged behind the ensemble methods. The Transformer model showed significant predictive failure, with an MRE of 0.047 and an MSE of 10346.86, which are 47 and 373 times higher than those of XGBoost, respectively, and the lowest \(R^2\) value (0.9533). This pronounced deviation contradicts the theoretical expectation of the Transformer’s ability to capture long-range dependencies and suggests that the standard attention mechanism may not be well suited for short input sequences \cite{vaswani2017attention, alammar2021non}.

The error distribution analysis further reveals key patterns: XGBoost demonstrated exceptional stability, with a prediction error standard deviation of 0.03 and an interquartile range (IQR) of 0.07. The Transformer model exhibited a consistent prediction bias across the entire transition region, indicating that its attention mechanism struggled to capture localized and abrupt transition features. The KAN model showed a right-skewed error distribution and a 22\% failure rate in early transition predictions (\(x < 700\)), likely due to its insufficient capacity to represent low-data-density regions—an issue commonly encountered in machine learning models when training data is sparse, as discussed in previous studies on machine learning and neural networks \cite{bishop2006pattern, smola1998kernel, he2016sparse}.

XGBoost demonstrated exceptional training efficiency, requiring only 0.9\,s, making it 11.6 times faster than CNN, which took 22.3\,s. This rapid training time is primarily attributed to XGBoost's tree-based structure, which benefits from parallelization, enabling faster convergence \cite{al2024comparative}. In contrast, LSTM exhibited the longest training time of 213.8\,s, reflecting the inefficiencies inherent in recurrent architectures, where each time step depends on the previous one, leading to higher computational costs \cite{cho2014learning}. Ensemble methods, such as XGBoost and Random Forest, require approximately 25\,MB of storage. Deep learning models, including DNN (0.7\,MB) and CNN (0.5\,MB), offer a reasonable balance between accuracy and storage efficiency. Transformer models (0.3\,MB) achieve extreme parameter compression but suffer from unacceptable prediction errors (MAE \(> 50\)), making them unsuitable for high-precision applications.

To further investigate the key features relied upon by the models during prediction, this study utilized the SHAP (Shapley Additive Explanations) method to assess the contribution of each feature in the predictive process for each model. As shown in Figure~\ref{fig:performance}d, for tree-based models, such as Random Forest and XGBoost, the feature importance distribution was relatively smooth and uniform, indicating that these models rely on the combination of features to effectively learn task-related patterns, with both time and spatial input dimensions contributing equally to the prediction \cite{liu2021controlburn}. In contrast, the LSTM model exhibited clear temporal dependencies, with the feature importance heatmap showing greater significance for earlier time steps, highlighting the LSTM’s advantage in capturing temporal relationships when handling time series data \cite{hochreiter1997long}. The Transformer model demonstrated significant spatial dependence, with the importance of the first monitoring point being notably higher than that of other positions, suggesting that its attention mechanism may focus on specific spatial locations \cite{guo2022transformer, li2021local}. The feature importance distributions for DNN, CNN, and KAN were more uniform, with no particular time step or spatial location standing out. This suggests that these models did not fully leverage local temporal information in time series data, aligning with their characteristics of global feature learning \cite{ren2021bidirectional}.

The results presented in Figure~\ref{fig:performance} provide a novel solution for the fine-grained prediction of laminar-turbulent transition with minimal pressure signal input. The limited input is crucial due to unpredictable environmental conditions and the complexity of precisely modeling the receptivity and nonlinear breakdown of disturbances. Compared to currently available methods, the proposed method accounts for varying receptivity results and the nonlinear development of perturbations, thereby accurately describing the physical transition process. The N-factor method does not consider receptivity and nonlinear effects, while the transitional turbulence modeling approach \cite{menter2006correlation, langtry2009correlation} does not attempt to model the physics of the transition process. However, its strength lies in its compatibility with modern computational fluid dynamics. The proposed method can be extended to accumulate larger datasets, identify the key common factors of transition, and integrate them into the computational fluid dynamics (CFD) framework, as well as measurement-based flow control.

\section*{DISCUSSION}\label{S4}

This study generated 3,000 datasets covering K-type, H-type, and O-type transition paths using NPSE. By combining wall-pressure signal features with machine learning models, it achieved high-precision, fine-grained prediction of boundary-layer transition locations, with an average relative error of approximately 0.001. This approach overcomes the limitations of traditional methods: unlike LST or PSE based $N$-factor method, it fully captures the entire nonlinear evolution of disturbances, including post-receptivity and breakdown; compared to direct numerical simulation (DNS), it significantly reduces computational costs while maintaining accuracy, offering a new paradigm for rapid transition prediction.

A key advantage lies in the physical completeness of the dataset. NPSE simulations encompass critical stages such as disturbance amplification, nonlinear interactions, and the formation of initial turbulence. This avoids the generalization deficiencies typical of purely data-driven models. Furthermore, the wall-pressure signals used are experimentally measurable, facilitating real-world applicability, while the efficiency of ensemble learning methods like XGBoost enables potential real-time prediction.

This work also points to future extensions. The current framework focuses on zero-pressure-gradient flat-plate boundary layers, excluding complex geometries and pressure-gradient variations commonly found in engineering scenarios (e.g., aircraft surfaces). In addition, the dataset does not cover extreme conditions such as hypersonic thermochemical nonequilibrium, which could play a vital role in model generalization for high-speed regimes. Future research may expand the dataset to include such complexities to enhance extrapolation capabilities, and explore multi-feature fusion with experimental data to construct a more comprehensive transition-signature library.

This work presents the first large-scale dataset for transitional flows under varying initial perturbations. It offers valuable insights into the prediction of other nonlinear systems (e.g., combustion instability, vascular turbulence) and may facilitate engineering applications of flow stability theory.

\section*{METHOD}\label{S3}

To accurately predict transition, it is crucial to create a comprehensive dataset that captures a broad range of initial conditions while also providing precise transition onset locations. Striking an optimal balance between efficiency and accuracy, the flow field under various initial conditions is computed by solving the NPSE. Quantitative comparisons with DNS results and benchmark cases show excellent agreement in both the nonlinear evolution of perturbations and the predicted transition onset points, thus validating the accuracy and applicability of the NPSE-based approach (see Appendix \ref{secA1}). Subsequently, wall-pressure signals at selected streamwise locations are extracted and mapped to transition characteristics, which are then used as input features for machine learning models to predict transition onset locations.

\subsubsection*{A comprehensive dataset}
\begin{figure}
\centering
\includegraphics[height=0.55\textwidth]{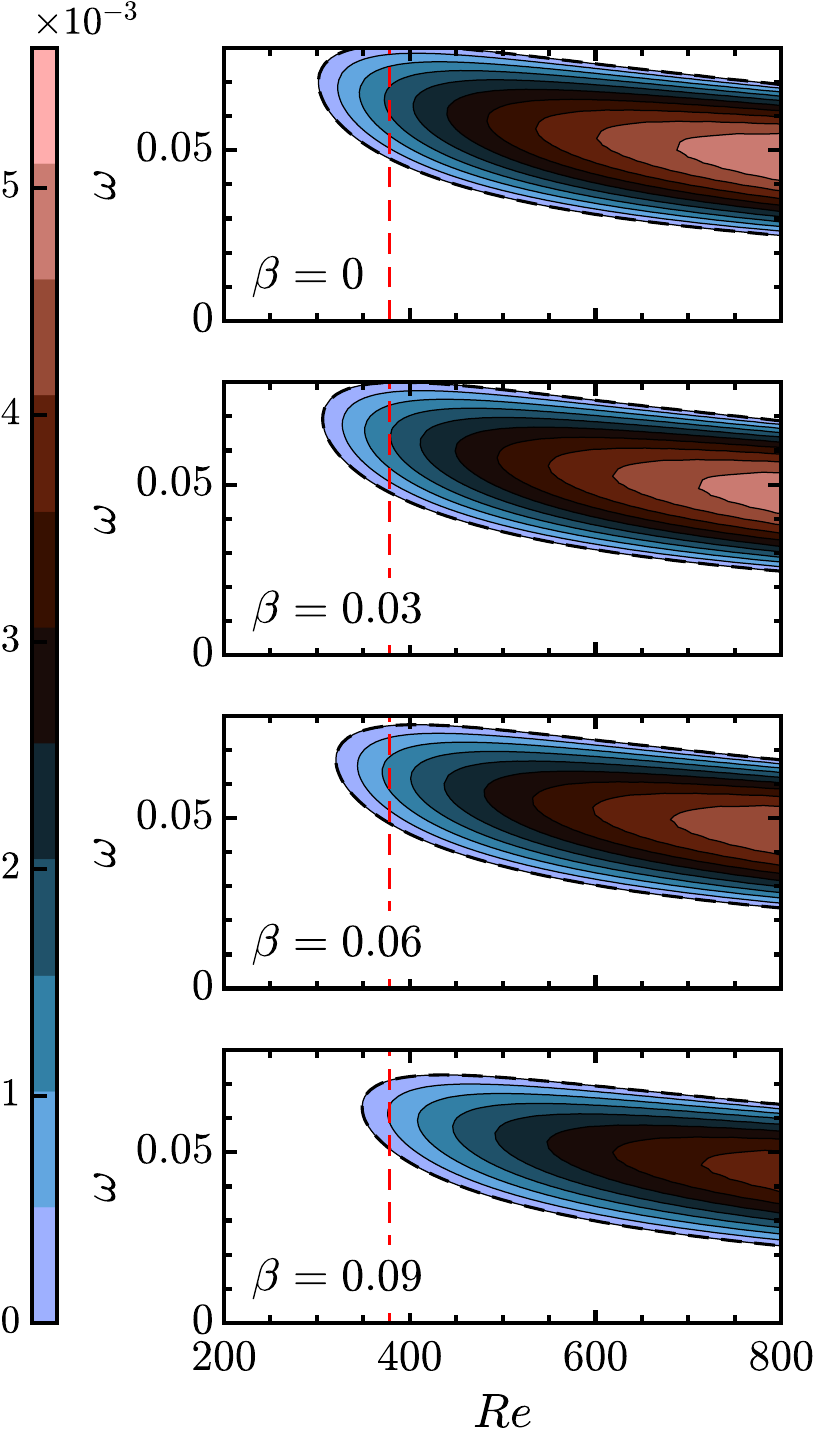}
\put(-133, 285){\large \bfseries\sffamily A}
\hspace{2mm}
\includegraphics[height=0.55\textwidth]{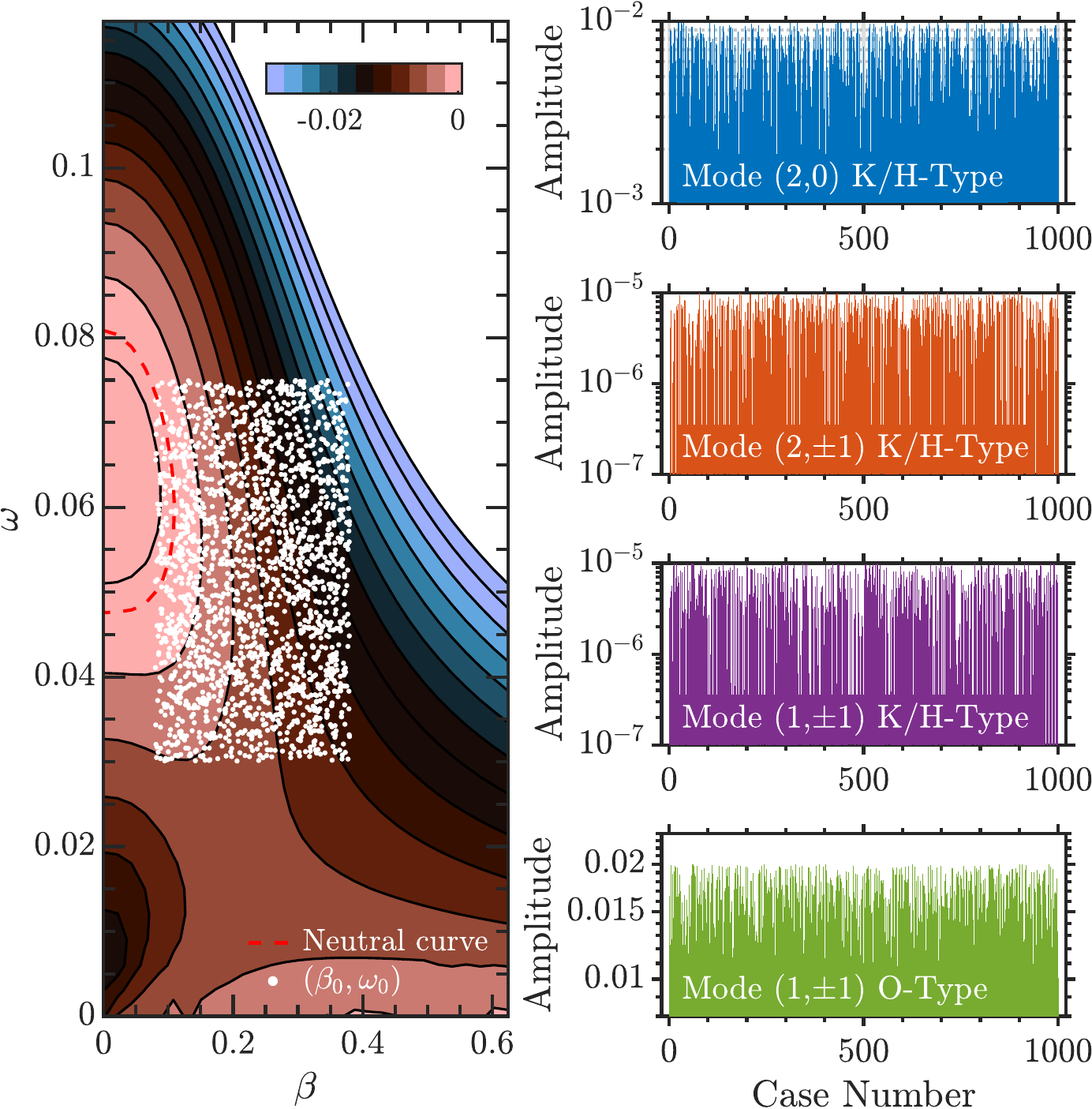}
\put(-270, 285){\large \bfseries\sffamily B}
\put(-115, 285){\large \bfseries\sffamily C}
\\[5mm]
\includegraphics[width=0.99\textwidth]{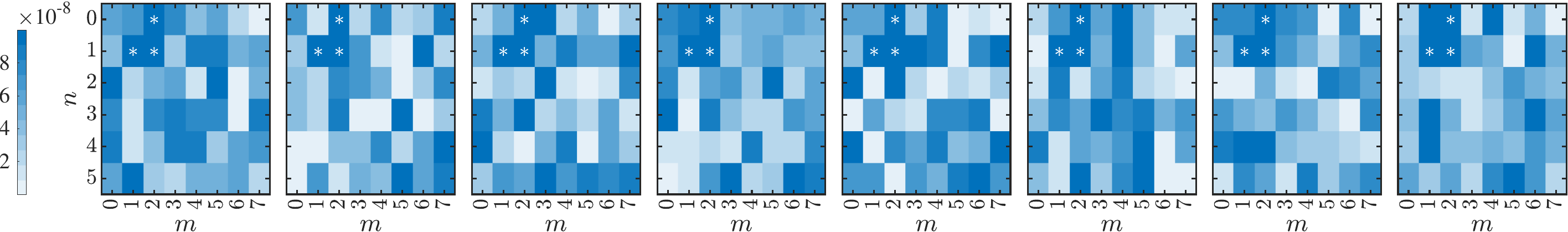}
\put(-485, 80){\large \bfseries\sffamily D}
\caption{
\textbf{\sffamily Initialized modes and parameters for dataset creation}.
\textbf{\sffamily A} Contours of the growth rate as functions of spanwise wavenumber \( \beta \) and Reynolds number \( Re \) at a freestream Mach number \( M = 0.01 \), with the neutral curve highlighted by the black dashed line. The red dashed line indicates the inlet Reynolds number \( Re = 378.3 \) used in the simulations.
\textbf{\sffamily B} Growth rate distribution in the \( \beta \)-\( \omega \) space (\( Re = 378.3 \)), where the red dashed line represents the neutral curve. The white dots denote the fundamental frequencies and wavenumbers \((\omega_0, \beta_0)\) of \textbf{\sffamily Dataset I}.
\textbf{\sffamily C} Amplitudes of initial modes for \textbf{\sffamily Dataset I}. From top to bottom: K-/H-type mode (2,0), (2, $\pm$1) and (1, $\pm$1),  O-type  mode (1, $\pm$1).
\textbf{\sffamily D} Amplitudes of the initial modes for \textbf{\sffamily Dataset II} as functions of \((m,n)\). Eight cases are shown. The asterisk (*) denotes the primary modes whose amplitudes exceed the maximum value shown in the color map. 
}
\label{fig:methods}
\end{figure}

We consider a flat-plate boundary-layer corresponding to a freestream Mach number of \( M = 0.01 \), where the laminar base flow is described by the self-similar solution. To construct a comprehensive dataset and initialize the NPSE simulations, two data groups (\textbf{\sffamily Dataset I} and \textbf{\sffamily Dataset II}) were defined, comprising a total of \( 3000 \) cases. The neutral stability curves shown in Figure~\ref{fig:methods} (a,b) help identify the range of relevant frequencies and spanwise wavenumbers used in the subsequent nonlinear computations \cite{mack1984boundary, malik1990numerical, van1951turbulent}. A random perturbation amplitude is assigned to each initialized mode (see Figure~\ref{fig:methods}c), with parameters selected based on the LST results. \textbf{\sffamily Dataset I} consists of \( 2000 \) cases, including \( 1000 \) cases for K- and H-type breakdown (initialized with modes \((2, 0)\), \((2, \pm 1)\), and \((1, \pm 1)\)), and \( 1000 \) cases for the O-type (initialized with mode \((1, \pm 1)\)). The fundamental frequency and wavenumber of these cases are also randomized as illustrated with the white dots in Figure~\ref{fig:methods}b. \textbf{\sffamily Dataset II} consists of \( 1000 \) cases, where perturbations are initialized by superimposing higher harmonics with lower amplitude onto the primary modes of K- and H-type transitions (see Figure~\ref{fig:methods}d), after fixing the fundamental frequency and wavenumber. These higher harmonics represent low-amplitude noise in Fourier space, which can be generated during the receptivity stage. The complete parameter set consists of \( 3000 \) unique combinations of initial conditions, which are summarized in Table \ref{tab:parameter_ranges}.

\renewcommand{\arraystretch}{1.2}
\begin{table}[ht]
\centering
\caption{Parameter ranges for disturbance configurations in {\bfseries\sffamily Dataset I} and {\bfseries\sffamily II}.}
\label{tab:parameter_ranges}
\begin{tabular}{clc}
\hline
\multirow{6}{*}{{\bfseries\sffamily Dataset I~}$\left\{\begin{array}{c} \\[2cm] \end{array}\right.$} & {Fundamental frequency $F_0=\omega_0/Re_0$} & $2 \times 10^{-5}$ to $9 \times 10^{-4}$ \\
& {Fundamental spanwise wavenumber $B_0=\beta_0/Re_0$} & $2 \times 10^{-4}$ to $7 \times 10^{-3}$ \\
& {Amplitude of mode $(2, 0) $(K-/H-type)} & $1 \times 10^{-3}$ to $1 \times 10^{-2}$ \\
& {Amplitudes of mode $(1, \pm1)$ \& $(2, \pm1)$ (K-/H-type)} & $1 \times 10^{-7}$ to $1 \times 10^{-5}$ \\
& {Amplitudes of mode $(1, \pm1)$ (O-type)} & $1 \times 10^{-2}$ to $2 \times 10^{-2}$ \\
& {Amplitudes of other harmonics $(m, \pm n)$} & --- \\
\hline
\multirow{6}{*}{{\bfseries\sffamily Dataset II}$\left\{\begin{array}{c} \\[2cm] \end{array}\right.$} & {Fundamental frequency $F_0=\omega_0/Re_0$} & $6.2 \times 10^{-5}$ \\
& {Fundamental spanwise wavenumber $B_0=\beta_0/Re_0$} & $3.3 \times 10^{-4}$ \\
& {Amplitude of mode $(2, 0) $(K-/H-type)} & $1 \times 10^{-3}$ to $1 \times 10^{-2}$ \\
& {Amplitudes of mode $(1, \pm1)$ \& $(2, \pm1)$ (K-/H-type)} & $1 \times 10^{-7}$ to $1 \times 10^{-5}$ \\
& {Amplitudes of mode $(1, \pm1)$ (O-type)} & --- \\
& {Amplitudes of other harmonics $(m, \pm n)$} & $1 \times 10^{-10}$ to $1 \times 10^{-7}$ \\
\hline
\end{tabular}
\end{table}

\subsubsection*{Feature extraction}
According to the ideal gas equation of state, the instantaneous pressure field is given by:
\begin{equation}
	p = \rho R_g T = (\rho_0 + \rho') R_g (T_0 + T') = \rho_0 R_g T_0 + \rho_0 R_g T' + \rho' R_g T_0 + \rho' R_g T',
\end{equation}
Here, the base flow pressure is defined as:
\begin{equation}
	p_0 = \rho_0 R_g T_0,
\end{equation}
and the pressure disturbance is expressed as:
\begin{equation}
	p' = \rho_0 R_g T' + \rho' R_g T_0 + \rho' R_g T'.
\end{equation}
Where \( R_g = {1}/{\gamma M^2} \), is the dimensionless specific gas constant. \( \gamma \) is the specific heat ratio, defined as the ratio of specific heats at constant pressure and volume \( \left( \gamma = {c_p}/{c_v} \right) \).
In practical measurements, pressure sensors are straightforward to deploy and provide reliable characterization of flow disturbances and boundary-layer transition \cite{mears2022unsteady}. Inspired by this, pressure data are extracted at several fixed streamwise locations, \(Re = [400, 417, 439, 461, 482, 502]\). These positions were selected because they are located well upstream of the transition onset, making them suitable for detecting early signs of transition. At each location, instantaneous pressure values are recorded over \(N_t = 50\) consecutive time steps, corresponding to one complete cycle of the signal. By using three to six wall-pressure signals, indicators of transition onset can be effectively identified.

In boundary-layer transition studies, the deviation of the $C_f$ from its laminar value is recognized as the onset of transition. The skin-friction coefficient $C_f$ is defined as:
\begin{equation}
	C_f = \frac{\tau_w}{0.5 \rho U_{\infty}^2}
\end{equation}
where $ \tau_w $ is the wall shear stress, given by

\begin{equation}
	\tau_w = \mu \left. \frac{\partial u}{\partial y} \right|_{y=0}.
\end{equation}
Here, $ \mu $ is the dynamic viscosity of the fluid, and $ y $ is the wall-normal distance from the wall. 
This criterion leverages the fact that in the laminar flow regime, \( C_f \) decreases gradually with increasing Reynolds number due to the increase in boundary-layer thickness. Upon transitioning to turbulent flow, the shear stress increases significantly. Therefore, the rise in \( C_f \) is defined as the transition onset, which initiates the increase in turbulent energy and the formation of turbulent structures. Consequently, in the data-processing stage, the local minimum of \( C_f(x) \) is identified and extracted to label the data of each case.

\subsubsection*{Nonlinear parabolized stability equations} 
NPSE are derived from the compressible Navier-Stokes (N-S) equations by retaining both non-parallel and nonlinear effects. The formulation is based on the assumption that the disturbance shape functions vary slowly in the streamwise direction, which allows for a parabolic marching approach. The equations are non-dimensionalized using the displacement thickness at the inlet, \( \delta_0^* \), freestream velocity, \( U_\infty^* \), freestream temperature, \( T_\infty^* \), and freestream density, \( \rho_\infty^* \). Here the superscript $()^*$ stands for dimensional quantities.
\begin{equation}\label{eq_non_dimensionalization}
\begin{array}{l}
    x,y,z = \dfrac{x^*,y^*,z^*}{\delta_0^*}, \quad
    t = \dfrac{t^* U_\infty^*}{\delta_0^*}, \quad
    u,v,w = \dfrac{u^*,v^*,w^*}{U_\infty^*}, \quad
    \rho = \dfrac{\rho^*}{\rho_\infty^*}, \quad
    T = \dfrac{T^*}{T_\infty^*}, \quad
    p = \dfrac{p^*}{\rho^*_\infty U_\infty^{*2}}, \quad
    \mu = \dfrac{\mu^*}{\mu^*_\infty}.
\end{array}
\end{equation}
The instantaneous flow field is decomposed as:
\begin{equation}
\boldsymbol{q} = \boldsymbol{Q}(x, y) + \tilde{\boldsymbol{q}}(x, y, z, t).
\end{equation}
where $\boldsymbol{Q}(x, y)$ is the baseflow, and $\tilde{\boldsymbol{q}}(x, y, z, t)$ denotes the perturbation. Both the baseflow $\boldsymbol{Q}(x, y)$ and the perturbed flow $\boldsymbol{q}$ satisfy the original N-S equations. Substituting into the governing equations and subtracting the base flow yields the perturbation system:
\begin{equation}\label{eq_stab}
\mathbf{\Gamma} \frac{\partial \tilde{\boldsymbol{q}}}{\partial t}
+ \mathbf{A} \frac{\partial \tilde{\boldsymbol{q}}}{\partial x}
+ \mathbf{B} \frac{\partial \tilde{\boldsymbol{q}}}{\partial y}
+ \mathbf{C} \frac{\partial \tilde{\boldsymbol{q}}}{\partial z}
+ \mathbf{D} \tilde{\boldsymbol{q}}
= \mathbf{V}_{xx} \frac{\partial^2 \tilde{\boldsymbol{q}}}{\partial x^2}
+ \mathbf{V}_{yy} \frac{\partial^2 \tilde{\boldsymbol{q}}}{\partial y^2}
+ \mathbf{V}_{zz} \frac{\partial^2 \tilde{\boldsymbol{q}}}{\partial z^2}
+ \mathbf{V}_{xy} \frac{\partial^2 \tilde{\boldsymbol{q}}}{\partial x \partial y}
+ \mathbf{V}_{yz} \frac{\partial^2 \tilde{\boldsymbol{q}}}{\partial y \partial z}
+ \mathbf{V}_{zx} \frac{\partial^2 \tilde{\boldsymbol{q}}}{\partial z \partial x}
+ \tilde{\boldsymbol{N}},
\end{equation}
where \( \mathbf{\Gamma}, \mathbf{A}, \mathbf{B}, \mathbf{C}, \mathbf{D}, \mathbf{V}_{xx}, \mathbf{V}_{yy}, \mathbf{V}_{zz}, \mathbf{V}_{xy}, \mathbf{V}_{yz}, \mathbf{V}_{zx} \) are \(5\times 5\) matrices, and \(\tilde{\boldsymbol{N}}\) is the nonlinear term, as described in previous works \cite{ren2019linear}. The perturbation is then expressed as a Fourier series in the spanwise and temporal directions:
\begin{equation}\label{eq_modal}
\tilde{\bm{q}} = \sum_{m=-M}^{M} \sum_{n=-N}^{N} \hat{\bm{q}}_{m,n}(x,y) \exp\left( i \int \alpha_{m,n}(x) \, dx + in\beta z - i m \omega t \right)
\end{equation}
where $\hat{\boldsymbol{q}}_{mn}(x, y)$ denotes the shape function of mode $(m,n)$, $\alpha_{mn}(x)$ is the complex streamwise wavenumber, $\beta$ is the spanwise wavenumber, and $\omega$ is the frequency. In the present study, the fundamental parameters are chosen as \( M = 7 \) and \( N = 5 \), which are large enough to provide accurate transition locations. Under the scale-separation assumption and by using the auxiliary conditions for the streamwise wavenumber \( \alpha_{mn} \), terms of order \( \mathcal{O}(1 / Re^2) \) or smaller are neglected and the system \eqref{eq_stab} reduces to a set of parabolic-type equations for each \( (m, n) \) mode:
\begin{equation}\label{eq_perturbation}
\widehat{\boldsymbol{A}} \frac{\partial \hat{\boldsymbol{q}}_{mn}}{\partial x} 
+ \widehat{\boldsymbol{B}} \frac{\partial \hat{\boldsymbol{q}}_{mn}}{\partial y} 
+ \widehat{\boldsymbol{D}} \hat{\boldsymbol{q}}_{mn} 
= \mathbf{V}_{yy} \frac{\partial^2 \hat{\boldsymbol{q}}_{mn}}{\partial y^2} 
+ \boldsymbol{F}_{mn} \exp\left( -i \int \alpha_{mn}(x)\, dx \right)
\end{equation}
where $\boldsymbol{F}_{mn}$ represents the nonlinear interaction among different modes. The coefficient matrices are given by:
\begin{equation}\label{eq_A}
\widehat{\mathbf{A}} = \mathbf{A} - 2 i \mathbf{V}_{xx} \alpha_{mn} - i n \beta \mathbf{V}_{xz},
\end{equation}
\begin{equation}\label{eq_B}
\widehat{\mathbf{B}} = \mathbf{B} - i \alpha_{mn} \mathbf{V}_{xy} - i n \beta \mathbf{V}_{yz},
\end{equation}
\begin{equation}\label{eq_D}
\widehat{\mathbf{D}} = - i m \omega \Gamma + i \alpha_{mn} \mathbf{A} + i n \beta \mathbf{C} + \mathbf{D} 
- i \frac{d\alpha_{mn}}{dx} \mathbf{V}_{xx} 
+ \alpha_{mn}^2 \mathbf{V}_{xx} 
+ (n \beta)^2 \mathbf{V}_{zz} 
+ \alpha_{mn} n \beta \mathbf{V}_{xz}.
\end{equation}

\subsubsection*{Machine learning models} 
\begin{figure}
\centering
\includegraphics[scale=0.7]{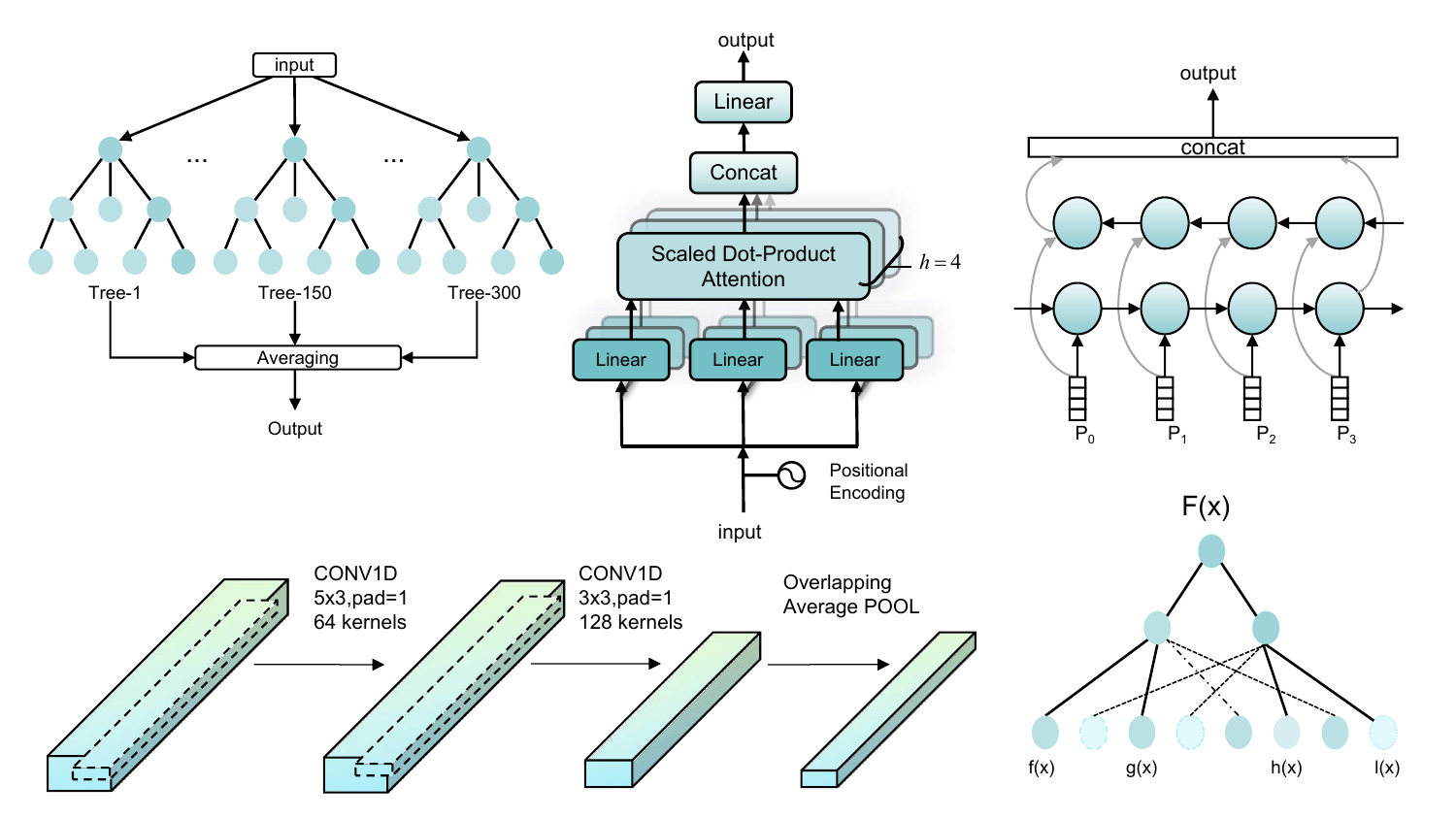} \put(-500,260){\large \bfseries\sffamily A}\put(-300,260){\large \bfseries\sffamily B}\put(-170,260){\large \bfseries\sffamily C}\put(-500,100){\large \bfseries\sffamily D}\put(-160,100){\large \bfseries\sffamily E}
\\
\caption{
\textbf{\sffamily Illustration of core architectures of various models}. 
\textbf{\sffamily A} Random Forest model consisting of 300 base learners (decision trees). Each decision tree is independently trained on different subsets of samples and features, and the final prediction is obtained by averaging the outputs of all trees.
\textbf{\sffamily B} Transformer multi-head self-attention mechanism. The input sequence is first combined with positional encoding to inject position information. Each attention head computes different attention weights via scaled dot-product attention, capturing global dependencies, with a total of four parallel heads ($h=4$). The outputs of all heads are concatenated along the channel dimension and passed through a subsequent linear layer to project into the target feature space.
\textbf{\sffamily C} BiLSTM network. The input consists of the pressure sequence at the current time point along with the preceding and following four time steps. Two LSTM subnetworks process the sequence in forward and backward directions, respectively, to capture temporal dependencies in both directions. The hidden states from both directions are concatenated and used for further regression modeling.
\textbf{\sffamily D} 1D CNN designed to extract local patterns in sequential data. The network includes two convolutional layers: the first layer uses 64 kernels with a kernel width of 5 (padding=1), and the second layer uses 128 kernels with a kernel width of 3 (padding=1), extracting features at different scales. The outputs of the convolutional layers are downsampled via overlapping average pooling to reduce sequence length and parameter count.
\textbf{\sffamily E} Kolmogorov-Arnold representation theorem. This theorem states that any multivariate continuous function $F(x)$ can be decomposed into a finite set of univariate continuous functions and additive combinations, enabling the approximation of complex high-dimensional nonlinear mappings by constructing a set of univariate functions (e.g., $f(x)$, $g(x)$, $h(x)$, $l(x)$) and their linear superpositions.}
\label{fig:ai}
\end{figure}
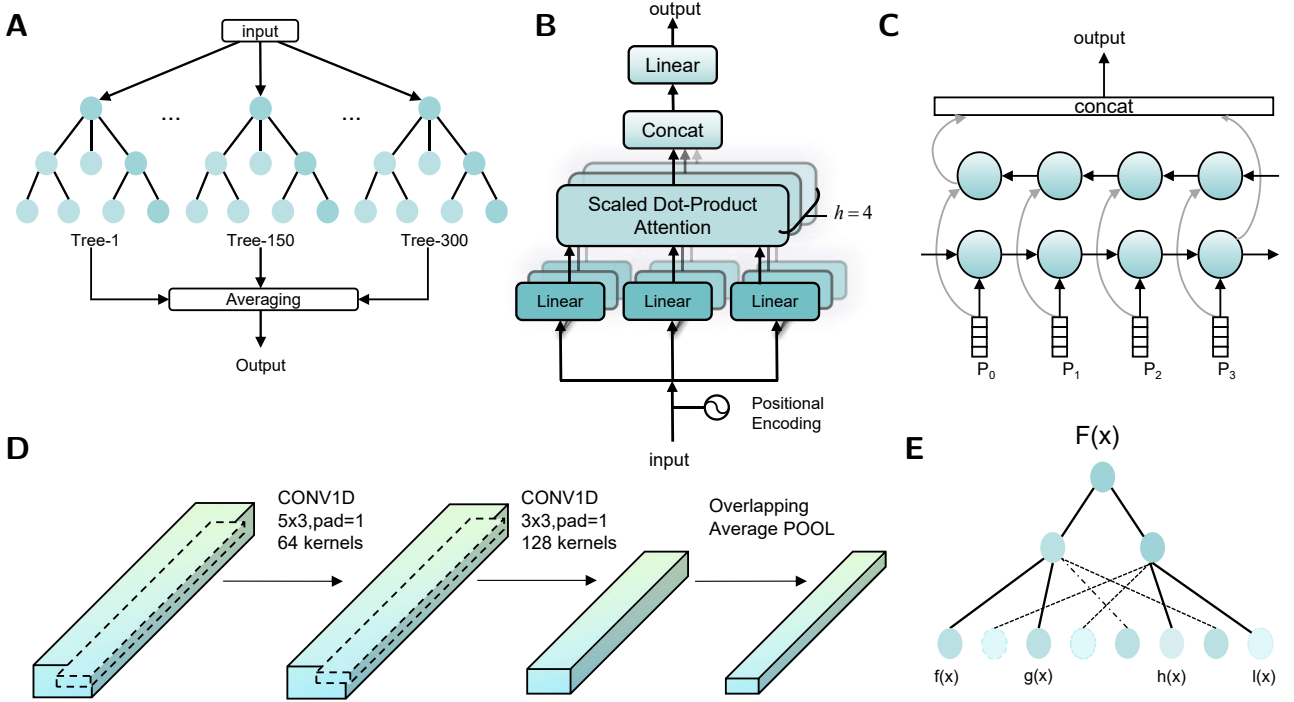

In this study, a machine learning-based time-series prediction framework is developed to predict the transition location in boundary-layer flow. The framework utilizes seven representative models, including deep neural networks, recurrent architectures, attention mechanisms, and ensemble learning methods. The core structures of these models are shown in figure \ref{fig:ai}. The input data for this framework consists of pressure signals extracted from the selected streamwise locations, as described above, while the output is the streamwise coordinate of the transition onset. These input-output configurations are consistent across all models, ensuring a fair performance comparison.

\begin{itemize}
\item{DNN:}  
A three-layer fully connected feedforward architecture is employed, taking a $N_t\times N_p$-dimensional input vector (corresponding to $N_t$ time steps recorded at $N_p$ spatial positions). \( N_p \) represents the number of pressure sensors used. The first hidden layer consists of 512 neurons with GELU activation and batch normalization. The second hidden layer reduces dimensionality to 256 neurons and incorporates a dropout rate of 0.3. The output layer is a single-node linear regression unit. The model is optimized using the Adam optimizer with an initial learning rate of $5 \times 10^{-4}$, a batch size of 64, and trained for 300 epochs.
\item{LSTM:}  
A two-layer bidirectional long short-term memory network is designed, with the first layer comprising 128 memory units (with \texttt{return\_sequences=True}), followed by a second layer of 64 memory units. Two fully connected layers with GELU activation follows, mapping the outputs to the prediction space. Training employs a dynamic learning rate strategy starting from 0.001, with a batch size of 32 and 200 epochs.
\item{CNN:}  
This architecture consists of two convolution pooling blocks: the first convolutional layer employs 64 channels with a kernel size of 5, followed by max pooling; the second convolutional layer uses 128 channels with a kernel size of 3; A fully connected layer. A global average pooling layer is used prior to the final linear regression output. The Adam optimizer (learning rate $5 \times 10^{-4}$) is adopted, with a batch size of 64 and 200 epochs.
\item{Transformer:}  
The input tensor with dimensions $(N_t, N_p)$ is processed with positional encoding and passed through a four-head multi-head self-attention mechanism with key dimension 16. Layer normalization is applied after each attention block. The output features are aggregated via global pooling before regression prediction. The model is trained using a learning rate of $1 \times 10^{-4}$, batch size 64, and 200 epochs.
\item{Random forest and XGBoost:}  
An ensemble of 300 regression trees is constructed for both RF and XGBoost using the CART algorithm, with a maximum tree depth of 12 for RF and a maximum depth of 6 for XGBoost. For RF, the variance reduction criterion is used for node splitting. The learning rate for XGBoost is set to 0.1, with a subsample ratio of 0.8 and regularization through a leaf weight shrinkage factor of 0.1. An early stopping mechanism is employed to prevent overfitting for both models.
\item{KAN:}  
Based on the Kolmogorov-Arnold theorem, a four-layer fully connected network is designed, taking a $N_t\times N_p$-dimensional input vector and progressively reducing it to 64 dimensions across successive layers. GELU activation, dropout, and layer normalization are incorporated throughout the network. Optimization is performed using the AdamW optimizer with a learning rate of 0.001 and a weight decay coefficient of 0.01.
\end{itemize}

All input data are standardized using Z-score normalization prior to model training. For recurrent and attention-based architectures, the input data are maintained in tensor form with dimensions \( (\text{sample size}, 50, N_p) \), whereas for non-sequential models, the data are reshaped into two-dimensional feature matrices of size \( (\text{sample size}, 50 \times N_p) \). All deep learning models implement an early stopping mechanism with a patience threshold of 20 epochs to prevent overfitting. The loss function for all neural network models is defined as the MSE. For ensemble models (RF and XGBoost), optimal hyperparameters are determined via grid search, and performance is assessed using the five-fold cross-validated coefficient of determination (\( R^2 \) score). In the context of transition prediction, the performance of these models is evaluated by comparing their ability to accurately predict the streamwise location of transition based on as few pressure signal inputs as possible.

\section*{Acknowledgements}
This work was supported by National Natural Science Foundation of China (Grant No.12372215), the Aeronautical Science Fund (Grant No. 2024M005072001) and the Alexander von Humboldt foundation.

\section*{Author Contributions}
W.C.:  Investigation, validation, writing - original draft and visualization. 
H.H.: Methodology, investigation, validation, investigation, writing - review and visualization. 
Y.X.: Framework development, writing - review. 
M.K.: Framework development, writing - review and J.R.'s AvH host.
H.T.: Resources, writing - review,  project administration, and funding acquisition.
J.R.: Con-ceptualization, methodology, resources, project administration, funding acquisition, writing - original draft and writing - review.

\section*{Materials and Correspondence}
Correspondence and requests for materials should be addressed to the corresponding authors (email: hhteng@bit.edu.cn; jie.ren@bit.edu.cn). The data supporting the findings of this study will be openly available upon publication of the manuscript.

\section*{Declaration of interests}
\noindent The authors report no conflict of interest.

\section*{Supplementary Materials}

\begin{figure}[h!]
\centering
\includegraphics[scale=0.26]{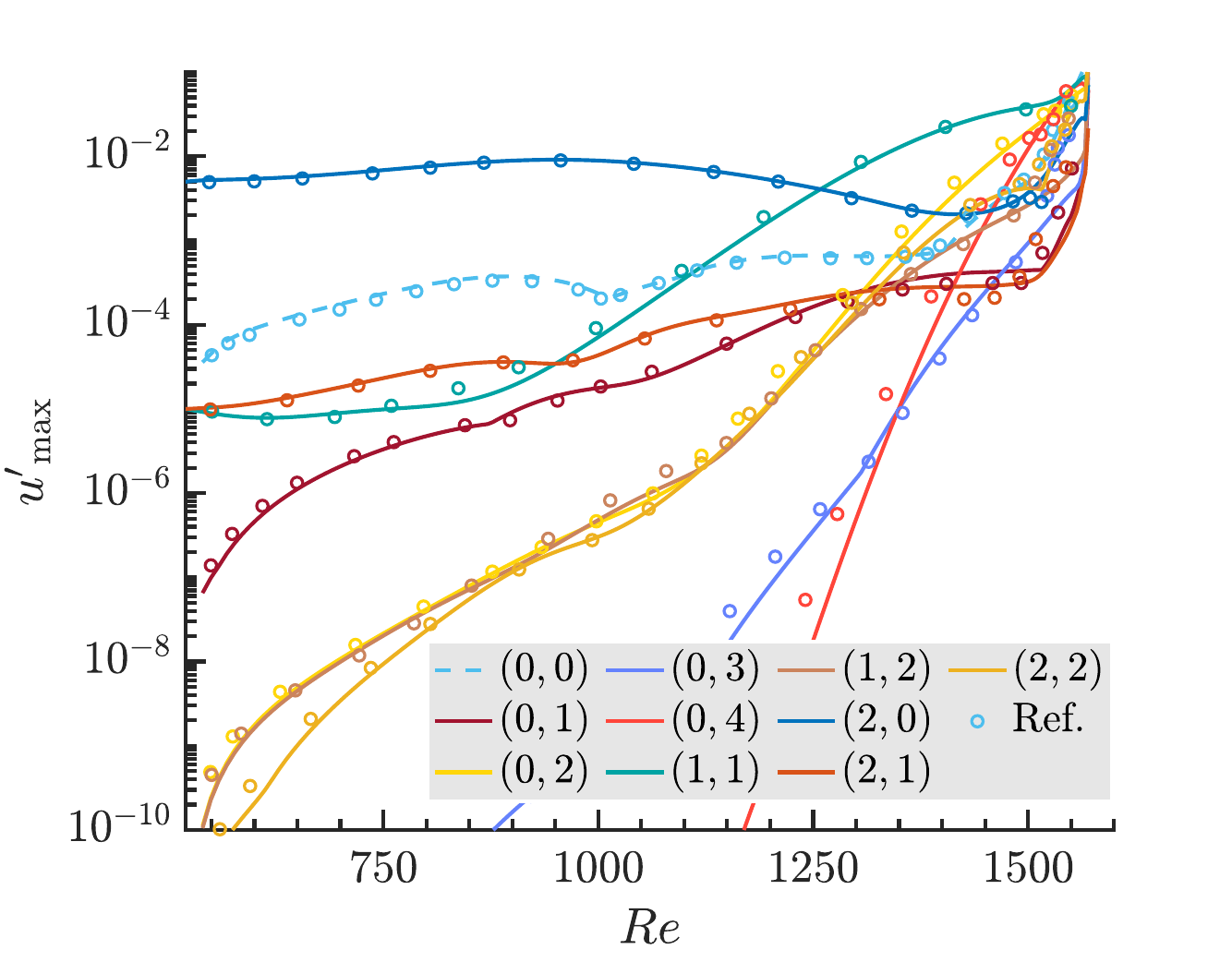} \put(-135, 115){\large \bfseries\sffamily A}
\includegraphics[scale=0.26]{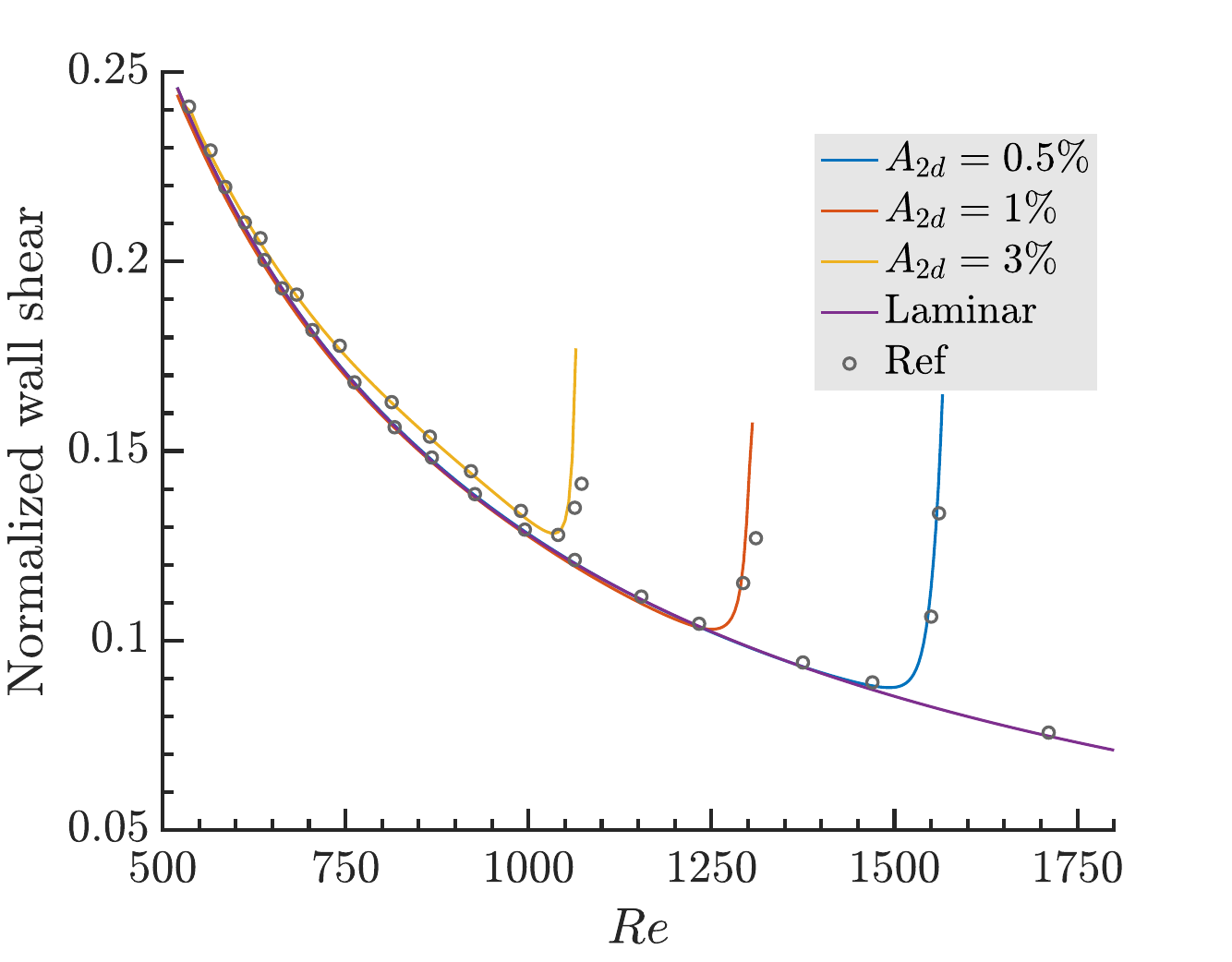} \put(-135, 115){\large \bfseries\sffamily B}
\includegraphics[scale=0.26]{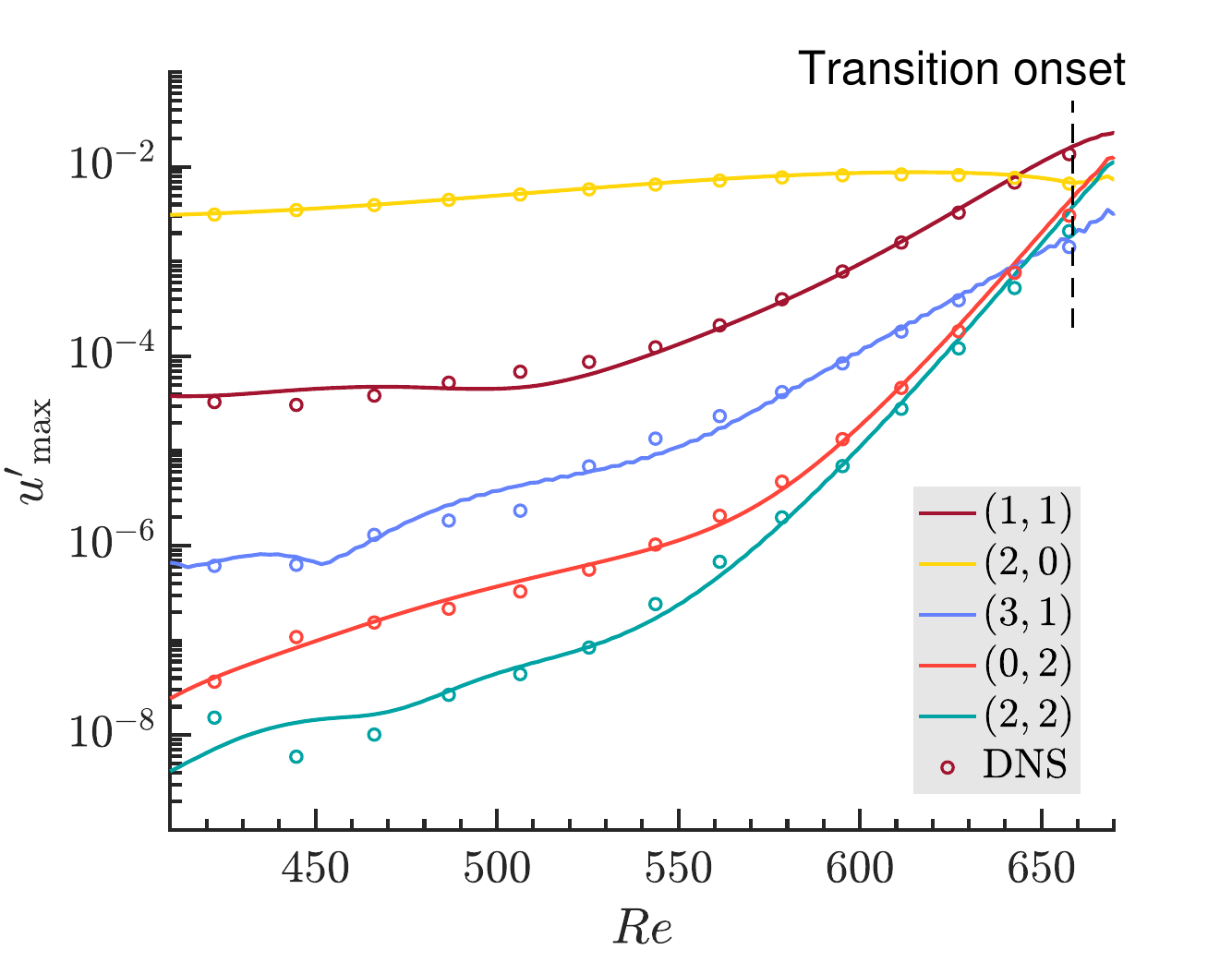} \put(-135, 115){\large \bfseries\sffamily C} \\
\includegraphics[scale=0.26]{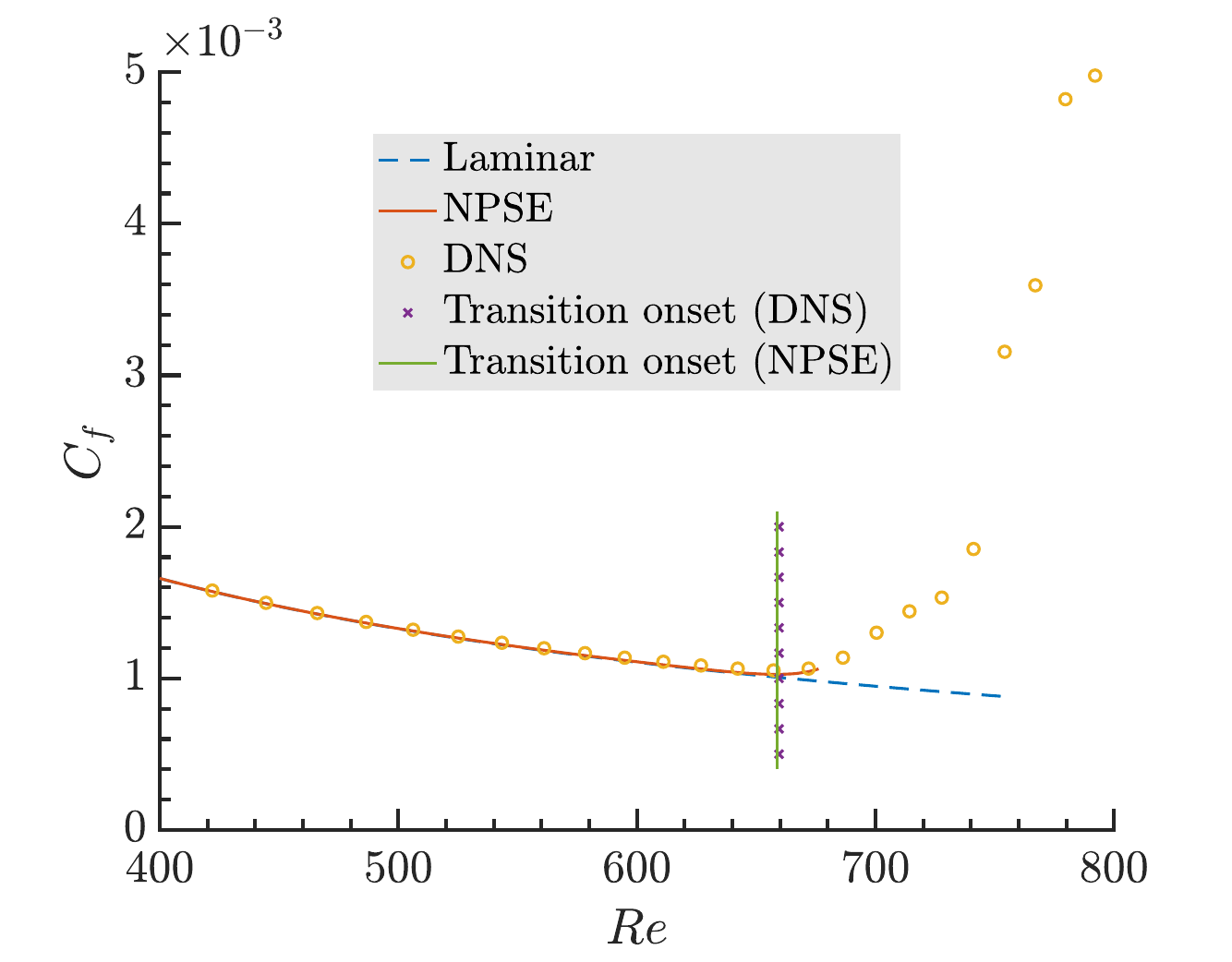} \put(-135, 110){\large \bfseries\sffamily D}
\includegraphics[scale=0.35]{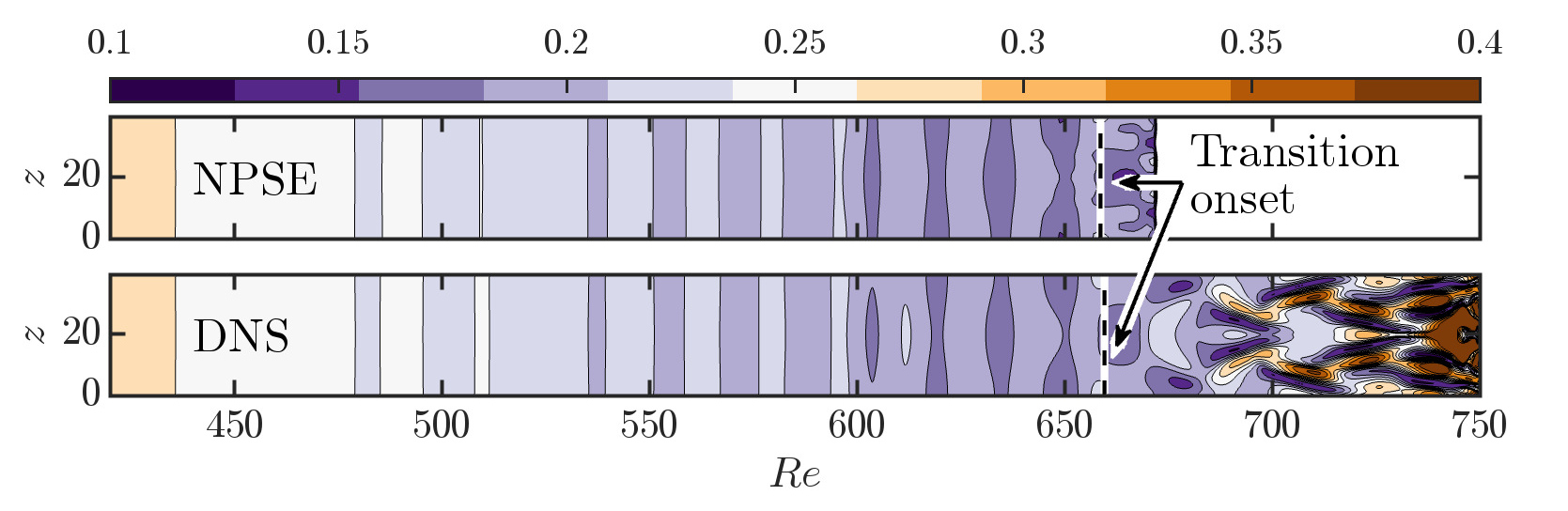} \put(-275, 90) {\large \bfseries\sffamily E}
\\
\includegraphics[clip, trim=12cm 2cm 15cm 8cm, scale=0.3]{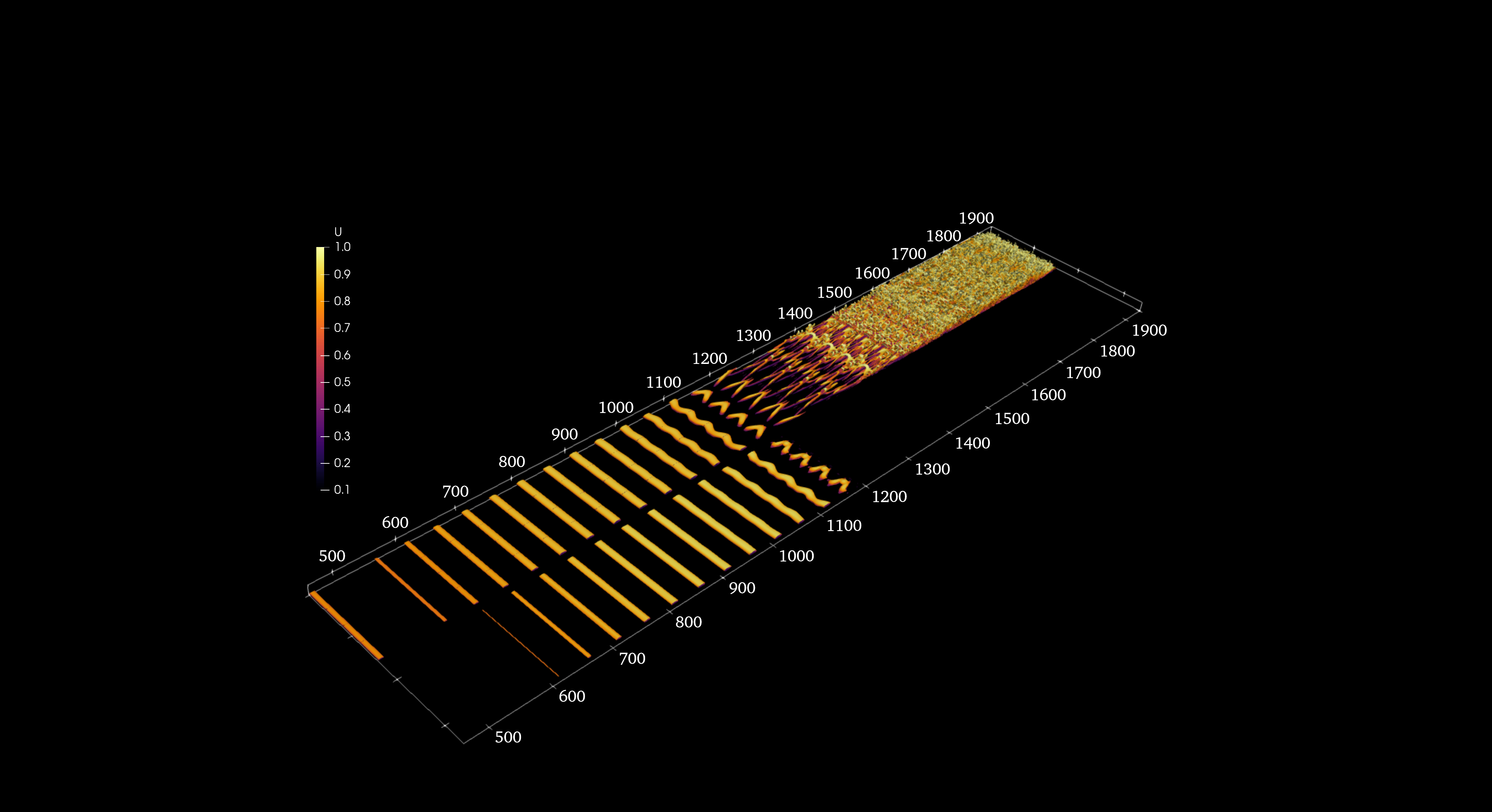} 
\put(-365, 225) {\White \large \bfseries\sffamily F}
\put(-335, 73) {\rotatebox{-45}{\White \large \bfseries DNS}}
\put(-310, 48) {\rotatebox{-45}{\White \large \bfseries NPSE}}
\caption{
\textbf{\sffamily Validation of NPSE Results with benchmark cases and DNS Data}. 
\textbf{\sffamily A} Comparison of perturbation amplitudes, $u_{\text{max}}'$ (lines), with data from \citet{chang1994oblique} (circles) as functions of Reynolds number $Re$. Each curve corresponds to a specific mode in Fourier space, represented by the double-spectral notation $(m,n)$. The initial amplitude of mode (2,0) is $A_{2d}=0.5\%$.
\textbf{\sffamily B} Comparison of the normalized wall shear stress versus $Re$ for different initial disturbance amplitudes: $A_{2d}=0.5\%$ (blue line), $1\%$ (red line), and $3\%$ (orange line). The circles indicate data from \citet{chang1994oblique}.
\textbf{\sffamily C} Comparison of NPSE results (lines) with DNS data (circles) for various modes, demonstrating the excellent accuracy of NPSE in Fourier space for a specific mode. 
\textbf{\sffamily D} Comparison of $C_f$ obtained by NPSE (red line) and DNS (circles). Both methods show excellent agreement in the transition onset, indicated by vertical lines. 
\textbf{\sffamily E} Comparison of the instantaneous streamwise velocity, $u$, at $y=0.9786$.
\textbf{\sffamily F} Comparison of the instantaneous flow structures using isosurfaces of the Q-criterion (Q=0.001).
}
\label{fig:validation}
\end{figure}

\appendix
\section{Validation of NPSE's applicability for building an encompassing dataset}\label{secA1}

The NPSE algorithm is validated in figure \ref{fig:validation} for both accuracy and suitability through comparisons with benchmark cases \cite{chang1994oblique} and DNS data. Its accuracy in resolving the flow field is demonstrated by the agreement in the Fourier amplitudes of disturbances, while its suitability is confirmed by the comparison of $C_f$ curves, which show that NPSE accurately predicts the transition onset location. 

In Figure~\ref{fig:validation}(a-b), different initial amplitudes of the two-dimensional mode (2,0) are considered: \(A_{2d} = 0.5\%\), \(1\%\), and \(3\%\), all with a frequency of \(F = 0.4 \times 10^{-4}\). Two pairs of symmetric three-dimensional modes, (2,\(\pm1\)) and (1,\(\pm1\)), are superimposed, with frequencies of \(0.4 \times 10^{-4}\) and \(0.2 \times 10^{-4}\), respectively, and a spanwise wavenumber of \(\beta/Re = \pm 0.96 \times 10^{-4}\). Their amplitudes are set to \(A_{3d} = 0.001\%\). The initial perturbations are obtained from LST at \(R = 520\). Figure~\ref{fig:validation}(c-f) present a comparison with DNS for an H-type transition, which includes mode (2,0) and modes (1,\(\pm1\)). The fundamental frequency is \(F = 62 \times 10^{-6}\), and the spanwise wavenumber is \(\beta = 3.3 \times 10^{-4}\). Initial conditions are specified at \(R = 380\), with amplitudes \(A_{2d} = 4.4\%\) and \(A_{3d} = 0.07\%\). These comparisons show excellent agreement.

\bibliography{ref}

\end{document}